\pdfoutput=1

\documentclass[11pt,a4paper]{article}

\usepackage{jheppub}
\usepackage{listings}
\usepackage{amsmath}
\usepackage{longtable}
\usepackage{slashed}
\usepackage{graphicx}
\usepackage{caption}
\usepackage{subcaption}
\usepackage{multirow}
\usepackage{amssymb}
\usepackage{mathrsfs}
\usepackage{bm}
\usepackage{xspace}
\usepackage{makecell}
\usepackage{booktabs}
\usepackage{xcolor}
\usepackage{IEEEtrantools}
\usepackage{verbatim}
\usepackage{fancyvrb}
\usepackage{tikzit}
\tikzstyle{dot}=[fill=black, draw=black, shape=coordinate, scale=0.2, font=tiny, tikzit draw=black, tikzit fill=black]
\tikzstyle{Sloped node}=[fill=none, draw=none, shape=circle, sloped, font=small, tikzit fill=magenta]
\tikzstyle{Label node}=[fill=none, draw=none, shape=circle, font=small]

\tikzstyle{Arrow}=[draw=black, ->]

\usepackage{axodraw2}
\newcommand{\eq}[1]{eq.~\eqref{eq:#1}}

\renewcommand{\sec}[1]{sec.~\ref{sec:#1}}

\newcommand{\fig}[1]{fig.~\ref{fig:#1}}

\newcommand{\abs}[1]{\lvert#1\rvert}

\newcommand{\nn}{\nonumber}

\newcommand{\df}{\mathrm{d}}

\newcommand{\Ga}{\Gamma}

\newcommand{\eps}{\epsilon}

\newcommand{\si}{\sigma}

\newcommand{\SUMMER}{\textsc{Summer}\xspace}
\newcommand{\XSUMMER}{\textsc{XSummer}\xspace}
\newcommand{\FORM}{\textsc{Form}\xspace}

\newcommand{\IBIS}{\textsc{IBIS}\xspace}
\newcommand{\SIGMA}{\textsc{Sigma}\xspace}

\allowdisplaybreaks[2]

\begin{document}

\title{\IBIS: Inverse BInomial sum Solver}

% Authors and affiliations
\author[a,b]{Paul A.J.W. van Hoegaerden}
\author[a,c]{Coenraad B. Marinissen}
\author[a,c]{Wouter J. Waalewijn}

% Affiliations
\affiliation[a]{Nikhef, Theory Group, Science Park 105, 1098 XG, Amsterdam, The Netherlands}
\affiliation[b]{ITF, Utrecht University, Leuvenlaan 4, 3584 CE Utrecht, The Netherlands}
\affiliation[c]{Institute for Theoretical Physics Amsterdam and Delta Institute for Theoretical Physics, University of Amsterdam, Science Park 904, 1098 XH Amsterdam, The Netherlands}

% E-mail addresses for corresponding authors
% \emailAdd{first.author@email.com}
% \emailAdd{second.author@email.com}

% Abstract
\abstract{In higher-loop calculations, Mellin-Barnes representations are used to simplify the denominators encountered in Feynman parameter integrals. The contour integrals of these representations yield sums over residues.
We develop an efficient method for certain classes of sums involving inverse binomials. Our results are expressed in terms of so-called $S$-sums, where the dependence on the upper limit of the sum is analytic. 
This is accomplished by deriving several new recursion relations, obtained from telescoping series and repeated synchronization.
We make our results available through \IBIS (``Inverse BInomial sum Solver''), a \FORM program to perform such inverse binomial sums. 
To highlight the efficiency of our code: sums up to weight 6 can be carried out in less than a second, and a comparison to the general-purpose \SIGMA and \textsc{EvaluateMultiSums} packages is included.
We show in examples how inverse binomial sums arise,  though some instances are beyond the cases studied here.
Our work thus provides a starting point for studying such sums and may open up new avenues in higher-loop calculations.

}

\date{\vspace{-1cm}}
\maketitle

%%%%%%%%%%%%%%%%%%%%%%
\section{Introduction} \label{sec:introduction}
%%%%%%%%%%%%%%%%%%%%%%

Collider physics has entered a precision era, driven by the demands of increasingly precise experimental measurements and enabled by substantial advances in theoretical calculations. 
QCD corrections are an essential component of these theoretical calculations, given the large size of the strong coupling, the fact that the LHC collides protons (QCD bound states) and that many analyses involve jets (QCD final states). Higher-order perturbative calculations give important corrections to predictions of the rates of scattering processes and substantially reduce the theory uncertainty.
The past two decades have brought significant advances, from automation of next-to-leading order (NLO) calculations~\cite{Ellis:2011cr} with a range of automated tools~\cite{Berger:2008sj,GoSam:2014iqq,Bevilacqua:2011xh,Cascioli:2011va,Badger:2012pg}, to next-to-next-to-leading order (NNLO) becoming the standard for $2 \to 2$ processes~\cite{Gehrmann-DeRidder:2019ibf,Czakon:2019tmo,Boughezal:2015dva,Gehrmann-DeRidder:2015wbt,Campbell:2016lzl,Chen:2019zmr,Czakon:2015owf,Catani:2019hip}. %
Even first NNLO results for $2 \to 3$ processes~\cite{Czakon:2021mjy} and next-to-next-to-next-to-leading order (N$^3$LO) results for $2 \to 1$ processes ~\cite{Anastasiou:2015vya,Dulat:2018bfe,Duhr:2020seh,Chen:2021vtu} have been achieved. 
Methodological advances in the calculation of two-loop amplitudes~\cite{Laporta:2000dsw, Gehrmann:1999as, Henn:2013pwa} and in the treatment of IR singular real radiation~\cite{Binoth:2004jv,Anastasiou:2003gr,Catani:2007vq,Gehrmann-DeRidder:2005btv,Currie:2013vh,Czakon:2010td,Boughezal:2011jf,Boughezal:2015dva,Boughezal:2015eha,Gaunt:2015pea,Somogyi:2008fc,DelDuca:2016ily}
were essential to this success, enabling predictions that are differential in the kinematics.
To obtain reliable predictions across the full kinematic range requires resummation, which can be achieved by matching to parton showers.

A specific ingredient in the evaluation of multi-loop integrals is Feynman parametrization, which however shifts computational challenges to difficult Feynman parameter integrals. These Feynman parameter integrals can be addressed by using Mellin–Barnes (MB) representations, rewriting them as contour integrals along the imaginary axis of the complex plane, involving ratios of Gamma functions. By closing the contour at infinity, these integrals are converted into infinite sums of residues via the residue theorem.
Subsequent operations, as outlined in \cite{Banik:2023rrz}, include resolving $1/\epsilon$ singularities of dimensionally ($d=4-2\eps$) regularized Feynman integrals~\cite{Smirnov:1999gc,Tausk:1999vh,Smirnov:2009up}, finding their analytic expressions in terms of hypergeometric functions~\cite{Kalmykov:2020cqz} or multiple polylogarithms~\cite{Vollinga:2004sn}, performing numerical integration~\cite{Dubovyk:2019krd}, etc. 
This strategy has been successfully applied in the analytical evaluation of Feynman integrals~\cite{Smirnov:2012gma,Dubovyk:2022obc}, and several computer codes have been developed to facilitate the construction of MB representations and the subsequent summation of residues~\cite{Czakon:2005rk,Gluza:2007rt,Smirnov:2009up,Ochman:2015fho}, which are summarized in~\cite{Belitsky:2022gba}. 

In this paper, we focus on the evaluation of MB integrals in terms of nested sums. Over the years, general classes of nested sums have been investigated, and their symbolic summation has led to the development of algorithms suitable for computer algebra systems. General-purpose tools for the symbolic evaluation of nested sums include \SIGMA and \textsc{EvaluateMultiSums}~\cite{SminaireLD,Ablinger:2010pb,Schneider:2013uan}. In most cases, the building blocks of these algorithms are the so-called $S$- and $Z$-sums, which generalize harmonic sums and (multiple) zeta values. Harmonic sums are the fundamental components of the \SUMMER~\cite{Vermaseren:1998uu} package in \FORM~\cite{Vermaseren:2000nd, Ruijl:2017dtg}. These sums frequently appear in single scale Feynman integrals~\cite{Blumlein:1998if}, for which a \texttt{Mathematica} implementation is also available~\cite{Ablinger:2014rba,Ablinger:2012ufz,Ablinger:2011te}. For Feynman integrals that depend on multiple scales, one can use the \FORM implementation \XSUMMER~\cite{Moch:2005uc}, or an implementation within the GiNaC framework~\cite{Bauer:2000cp,Vollinga:2005pk}. The availability of both the \SUMMER and \XSUMMER packages, coupled with \FORM's exceptional speed and efficiency in managing large expressions, are the primary reasons we selected \FORM for our implementation.

A particularly interesting class of sums encountered in multi-loop calculations using the MB representation takes the following form
\begin{equation}
\sim \sum_{j=1}^{n-1} \frac{1}{\binom{n}{j}}... \,\,,
\end{equation}
where the dots denote additional terms, e.g.~harmonic sums. In this paper, we refer to such sums as {\em inverse binomial sums}. In particular, the classical identity
\begin{align}
    \sum_{j=0}^n \frac{1}{\binom{n}{j}} = \frac{n+1}{2^{n+1}} \sum_{j=1}^{n+1} \frac{2^j}{j} \,,
\end{align}
which is the simplest example of the sums we consider here, has been rigorously proven in earlier works~\cite{gould1972combinatorial,Rockett01121981,pla1997sum,Trif01022000,SURY1993351}. Further research has explored more intricate identities~\cite{MANSOUR2002196,Sury2004,sym13112002}, including those incorporating harmonic sums~\cite{2014arXiv1411.6916J,2018arXiv180603022B}, as well as algorithmic methods for inverse-binomial sums~\cite{Schneider:2013uan}. In this paper, we develop an efficient method for rewriting the basis set of inverse binomial sums in tab.~\ref{table:possible sums} in terms of so-called $S$-sums, which we make available through an accompanying \FORM package called \IBIS (``Inverse BInomial sum Solver''). The parameters defining the structure of these inverse binomial sums are summarized in tab.~\ref{table:possible constants}. Specifically, tab.~\ref{table:possible constants} provides the numerical or symbolic ranges and constraints for these variables.

\begin{table}[t]
\centering
\renewcommand{\arraystretch}{2}
%\resizebox{\textwidth}{!}{
\begin{tabular}{ |c|c| }
\hline
   \multicolumn{2}{|c|}{Fixed sign} \\
\hline
  \makecell{0 harmonic \\ sums} & $\sum_{j=\text{max}\{1,-c+1\}}^{n-1} \frac{j!(n-j)!}{n!}\frac{1}{(j+c)^k}$  \\  
  \makecell{1 harmonic \\ sum} &  \makecell{$\sum_{j=\text{max}\{1,-c+1\}}^{n-1} \frac{j!(n-j)!}{n!}\frac{1}{(j+c)^k}S_{p_1,...,p_s}(n-j)$ \\ and $\sum_{j=\text{max}\{1,-c+1\}}^{n-1} \frac{j!(n-j)!}{n!}\frac{1}{(j+c)^k}S_{q_1,...,q_r}(j)$}  \\
  \makecell{2 harmonic \\ sums} & $\sum_{j=\text{max}\{1,-c+1\}}^{n-1} \frac{j!(n-j)!}{n!}\frac{1}{(j+c)^k}S_{p_1,...,p_s}(n-j)S_{q_1,...,q_r}(j)$  \\
 \hline
  \multicolumn{2}{|c|}{Sign altering} \\
 \hline
  \makecell{0 harmonic \\ sums} & $\sum_{j=\text{max}\{1,-c+1\}}^{n-1} \frac{j!(n-j)!}{n!}\frac{(-1)^j}{(j+c)^k}$ \\
  \makecell{1 harmonic \\ sum} & \makecell{$\sum_{j=\text{max}\{1,-c+1\}}^{n-1} \frac{j!(n-j)!}{n!}\frac{(-1)^j}{(j+c)^k}S_{p_1,...,p_s}(n-j)$ \\ and $\sum_{j=\text{max}\{1,-c+1\}}^{n-1} \frac{j!(n-j)!}{n!}\frac{(-1)^j}{(j+c)^k}S_{q_1,...,q_r}(j)$} \\
  \makecell{2 harmonic \\ sums} & $\sum_{j=\text{max}\{1,-c+1\}}^{n-1} \frac{j!(n-j)!}{n!}\frac{(-1)^j}{(j+c)^k}S_{p_1,...,p_s}(n-j)S_{q_1,...,q_r}(j)$ \\
  \hline
\end{tabular}
%}
\caption{The basis set of inverse binomial sums solvable using \IBIS. The $S$-sums appearing here are defined in \eq{harmonic}.
Using procedures provided by \IBIS and \SUMMER a larger set can be solved, as discussed in the \IBIS manual.}
\label{table:possible sums}
\end{table}

\begin{table}[t!]
\centering
\begin{tabular}{ |c|c| }
\hline
  Variable & Set of possible values \\
\hline
  $c$ & $\mathbb{Z}$ \\  
  $k$ & $\mathbb{N}_{0}$ \\
  $p_i,q_j$ & $\mathbb{Z}_{\neq 0}$ \\
  $s$,$r$ & $\mathbb{N}_{0}$ \\
  $n$ & \makecell{Symbolic: $n$ \\ Numerical: $\mathbb{N}_{>1}$} \\
 \hline
\end{tabular}
\caption{Possible values for the parameters  in tab.~\ref{table:possible sums} that are supported by \IBIS.}
\label{table:possible constants}
\end{table}

To illustrate the relevance of these sums, we provide a concrete example in app.~\ref{app:example}, which leads to many nested sums. Let us highlight a few examples of some terms that appear in this calculation which include inverse binomial sums. First of all, we have the following double sum  
\begin{align}
\sum_{k_1=0}^\infty \sum_{k_2=0}^\infty 
&(-1)^{k_1}\frac{k_1!k_2!}{(k_1+k_2+2)!}\frac{S_1(k_1+1)S_1(k_2)}{(k_1+k_2+1)(k_1+1)}\nn\\
&= -\sum_{k_2=1}^\infty \frac{1}{k_2(k_2+1)}\sum_{k_1=1}^{k_2-1}
(-1)^{k_1}\frac{k_1!(k_2-k_1)!}{k_2!}\frac{S_1(k_1)S_1(k_2-k_1)}{k_1^2}\,.\label{eq:exampleDouble}
\end{align}
which is of the last type as given in tab.~\ref{table:possible sums}. Here $S_1$ is a harmonic sum, defined in \eq{harmonic}. As a second example, consider the triple sum
\begin{align}
\sum_{k_1=0}^\infty \sum_{k_2=0}^\infty \sum_{k_3=0}^\infty
&(-1)^{k_1}
\frac{k_1! k_2!k_3!}{(k_1+k_2+k_3+2)!}
\frac{1}{(k_1+1)(k_1+k_2+2)(k_1+k_2+1)(k_3+1)}\nn\\
&= -\sum_{k_3=2}^\infty
\sum_{k_2=1}^{k_3-1} 
\frac{k_2! (k_3-k_2)!}{k_3!}
\frac{1}{(k_2+1)k_2(k_3-k_2)^2}
\sum_{k_1=1}^{k_2}
(-1)^{k_1}
\frac{k_1! (k_2-k_1)!}{k_2!}
\frac{1}{k_1^2}\label{eq:exampleTriple}
\,,
\end{align}
where both the sum over $k_1$ and $k_2$ include the inverse binomial factor. 
A final sum we want to highlight is given by
\begin{equation}\label{eq:finalexampleIBIS}
\sum_{k_2=3}^\infty \frac{1}{(k_2+3)(k_2+2)(k_2+1)}
\sum_{k_3=2}^{k_2-1} \frac{k_3!(k_2-k_3)!}{k_2!}
\sum_{k_1=1}^{k_3-1} \binom{k_1+k_2-k_3+1}{k_1}\frac{S_2(k_3-k_1)}{k_3-k_1+1} \,,
\end{equation}
with $S_2$ a harmonic sum as well.
In this expression, one must first evaluate the sum over $k_1$ that includes a binomial coefficient before the inverse binomial sum in $k_3$ can be computed, which is not yet supported by \IBIS. Such binomial sums have been studied before~\cite{Ablinger:2014wca,Ablinger:2014bra,Weinzierl:2004bn,Jegerlehner:2002em,Fleischer:1998nb}, and can be handled using \SIGMA~\cite{SminaireLD} and \textsc{EvaluateMultiSums}~\cite{Ablinger:2010pb,Schneider:2013uan}. These are general-purpose summation tools that can handle a much broader range of sums, but are therefore not as fast for higher-weight inverse binomial sums, as we will show. Just using these summation techniques was sufficient to solve certain Feynman integrals~\cite{Ablinger:2012qm,Ablinger:2015tua,Ablinger:2018brx}.
Specific classes of infinite inverse binomial sums involving harmonic sums were solved in refs.~\cite{Davydychev:2003mv, Kalmykov:2007dk}.

In \sec{usage}, we describe how \IBIS can be used to solve these example inverse binomial sums. In addition to the \IBIS code, a detailed manual and more concrete examples are available at~\href{https://github.com/cbmarini/IBIS}{IBIS}.

This paper is structured as follows. In the next section, we introduce the notation of the basic building blocks of nested sums. In sec.~\ref{sec:Recursionsec}, we explain how a recursion can be set up to express the inverse binomial sums in terms of $S$-sums. 
Sec.~\ref{sec:usage} describes the usage of \IBIS, its input and output and relevant \FORM notation for these sums, and how \IBIS can be used in the examples above. We also highlight the speed of our program, comparing to \textsc{EvaluateMultiSums}. We conclude and discuss directions for further research in sec.~\ref{sec:conclusion}.

%%%%%%%%%%%%%%%%%%%%%%%%%%%%%%%%%%%%%%%%%%%%%%%%%%%
\section{Definitions and properties of nested sums}
\label{sec:nestedSums}
%%%%%%%%%%%%%%%%%%%%%%%%%%%%%%%%%%%%%%%%%%%%%%%%%%%
As explained in the introduction, \IBIS computes inverse binomial sums in terms of $S$-sums. In this section, we provide an introduction to $S$-sums and discuss their key properties. Although the program primarily operates with $S$-sums, we will also provide the definition of $Z$-sums and explain their connection to $S$-sums. Additionally, we will highlight essential and useful procedures from \XSUMMER that can assist the user in preparing their inverse binomial sums as input for \IBIS. In the rest of this article, it will be implied that $p_1,...,p_k \in \mathbb{Z} $, $x_1,...,x_k \in \mathbb{R}_{\geq 0}$ and $n \in \mathbb{N}$. \\

\noindent \textbf{Definition of $S$-sums and $Z$-sums} \\
We define $S$-sums as follows~\cite{Moch:2001zr,Ablinger:2013cf}
\begin{equation}
S(n; p_1, \dots, p_k; x_1, \dots, x_k) = \sum_{n \geq i_1 \geq i_2 \geq \dots \geq i_k \geq 1} \frac{(\text{sign}(p_1))^{i_1}x_1^{i_1}}{i_1^{\abs{p_1}}} \dots \frac{(\text{sign}(p_k))^{i_k}x_k^{i_k}}{i_k^{\abs{p_k}}} \, , \label{eq:$S$-sumdef}
\end{equation}
where 
\begin{align}
    \text{sign}(x) = 
    \begin{cases}
    1 & \text{if} \quad x \geq 0 \,,\\
    -1 & \text{if} \quad x < 0 \,.
    \end{cases}
\end{align}
In a similar way, $Z$-sums can be defined by
\begin{equation}
Z(n;p_1, \dots, p_k; x_1, \dots, x_k) = \sum_{n \geq i_1 > i_2 > \dots > i_k > 0} \frac{(\text{sign}(p_1))^{i_1}x_1^{i_1}}{i_1^{\abs{p_1}}} \dots \frac{(\text{sign}(p_k))^{i_k}x_k^{i_k}}{i_k^{\abs{p_k}}} \, .
\end{equation}
For both $S$-sums and $Z$-sums we refer to $n$ as the \textit{argument} and $p_i$ as the \textit{indices} of the sums. \\

\noindent \textbf{Relation between $S$-sums and $Z$-sums} \\$S$-sums are closely related to $Z$-sums, the difference being the upper summation boundary for the nested sums: $i_m\geq i_{m+1}$ for $S$-sums, $i_m>i_{m+1}$ for $Z$-sums. They can be expressed in terms of each other, as discussed below. The reason both $Z$-sums and $S$-sums are used, is because some properties are more naturally expressed in terms of $Z$-sums while others are better represented using $S$-sums. In this work we use $S$-sums, as motivated in sec.~\ref{sec:telescoping}. 

An $S$-sum can be converted into $Z$-sums, by recursively using
\begin{align}
    S(n; p_1, &\dots, p_k; x_1, \dots, x_k) \nn \\
    &= \sum_{i_1=1}^n \frac{\text{sign}(p_1)^{i_1}x_1^{i_1}}{i_1^{\abs{p_1}}} \sum_{i_2=1}^{i_1-1} \frac{\text{sign}(p_2)^{i_2}x_2^{i_2}}{i_2^{\abs{p_2}}} S(i_2; p_3, \dots, p_k; x_3, \dots, x_k) \nonumber \\
    &\quad + S(n; \text{sign}(p_1\cdot p_2)(\abs{p_1} + \abs{p_2}), p_3, \dots, p_k; x_1 \cdot x_2, x_3, \dots, x_k) \,.
\end{align}
The key point is that the upper bound on $i_2$ in the first term on the right-hand side is $i_1-1$ instead of $i_1$, and the second term has one less index.
Similarly, one can convert a $Z$-sum into $S$-sums by recursive application of:
\begin{align}
    Z(n; p_1, &\dots, p_k; x_1, \dots, x_k) \nn \\
    &= \sum_{i_1=1}^n \frac{\text{sign}(p_1)^{i_1}x_1^{i_1}}{i_1^{\abs{p_1}}} \sum_{i_2=1}^{i_1} \frac{\text{sign}(p_2)^{i_2}x_2^{i_2}}{i_2^{\abs{p_2}}} Z(i_2 - 1; p_3, \dots, p_k; x_3, \dots, x_k) \nonumber \\
    &\quad - Z(n;\text{sign}(p_1 \cdot p_2) (\abs{p_1} + \abs{p_2}), p_3, \dots, p_k; x_1 \cdot x_2, x_3, \dots, x_k) \,.
\end{align}
The process of converting an $S$-sum into $Z$-sums can be easily programmed into \FORM and has been implemented in \XSUMMER. \\

\noindent \textbf{Reducing products of $S$-sums} \\
Furthermore, $S$-sums and $Z$-sums obey an algebra, allowing one to rewrite a product of two $S$-sums with the same argument in terms of a single $S$-sum. This can be done by recursive application of the following equation:
\begin{align}
    &S(n; p_1, \dots, p_k; x_1, \dots, x_k) \cdot S(n; p_1', \dots, p_l'; x_1', \dots, x_l') \nn \\
    &= \sum_{i_1=1}^n \frac{\text{sign}(p_1)^{i_1}x_1^{i_1}}{i_1^{\abs{p_1}}} S(i_1; p_2, \dots, p_k; x_2, \dots, x_k) S(i_1; p_1', \dots, p_l'; x_1', \dots, x_l') \nn \\
    &\quad + \sum_{i_2=1}^n \frac{\text{sign}(p_1')^{i_2}x_1'^{i_2}}{i_2^{\abs{p_1'}}} S(i_2; p_1, \dots, p_k; x_1, \dots, x_k) S(i_2; p_2', \dots, p_l'; x_2', \dots, x_l') \nonumber \\
    &\quad - \sum_{i=1}^n \frac{\text{sign}(p_1\cdot p_1')^{i}(x_1 \cdot x_1')^i}{i^{\abs{p_1} + \abs{p_1'}}} S(i; p_2, \dots, p_k; x_2, \dots, x_k) S(i; p_2', \dots, p_l'; x_2', \dots, x_l') \,. \label{eq:$S$-sum algebra}
\end{align}
Note that in the terms on the right-hand side, one or both of the $S$-sums have fewer indices than the product of $S$-sums on the left-hand side. This ensures that the recursion will ultimately terminate, resulting in a sum consisting solely of individual $S$-sums. The underlying algebraic structure in \eq{$S$-sum algebra} is a Hopf algebra, being realized as a quasi-shuffle algebra here, see e.g. \cite{Hoffman:2004bf, Weinzierl:2006qs}.
\begin{figure}[!t]
\centering
\resizebox{\textwidth}{!}{\begin{tikzpicture}[scale=0.5]
	\begin{pgfonlayer}{nodelayer}
		\node [style=none, label={below:$i$}] (16) at (-5, -2) {};
		\node [style=none] (17) at (-7.5, -2) {};
		\node [style=none] (18) at (-10, -2) {};
		\node [style=none] (19) at (-10, 0.5) {};
		\node [style=none, label={left:$j$}] (20) at (-10, 3) {};
		\node [style=none] (21) at (-3.5, 0.5) {=};
		\node [style=none] (22) at (-2, -2) {};
		\node [style=none] (23) at (-2, 0.5) {};
		\node [style=none, label={left:$j$}] (24) at (-2, 3) {};
		\node [style=none] (25) at (0.5, -2) {};
		\node [style=none, label={below:$i$}] (26) at (3, -2) {};
		\node [style=none] (27) at (4.5, 0.5) {+};
		\node [style=none] (28) at (6, -2) {};
		\node [style=none] (29) at (6, 0.5) {};
		\node [style=none, label={left:$j$}] (30) at (6, 3) {};
		\node [style=none] (31) at (8.5, -2) {};
		\node [style=none, label={below:$i$}] (32) at (11, -2) {};
		\node [style=none] (33) at (12.5, 0.5) {-};
		\node [style=none] (34) at (14, -2) {};
		\node [style=none] (35) at (14, 0.5) {};
		\node [style=none, label={left:$j$}] (36) at (14, 3) {};
		\node [style=none] (37) at (16.5, -2) {};
		\node [style=none, label={below:$i$}] (38) at (19, -2) {};
		\fill (-9, -1) circle (6pt);
		\fill (-9, 0) circle (6pt);
		\fill (-8, -1) circle (6pt);
        \fill (-7, -1) circle (6pt);
        \fill (-6, -1) circle (6pt);
        \fill (-8, 0) circle (6pt);
        \fill (-7, 0) circle (6pt);
        \fill (-6, 0) circle (6pt);
        \fill (-9, 1) circle (6pt);
        \fill (-8, 1) circle (6pt);
        \fill (-7, 1) circle (6pt);
        \fill (-6, 1) circle (6pt);
        \fill (-9, 2) circle (6pt);
        \fill (-8, 2) circle (6pt);
        \fill (-7, 2) circle (6pt);
        \fill (-6, 2) circle (6pt);
        \fill (16, 0) circle (6pt);
        \fill (15, -1) circle (6pt);
        \fill (17, 1) circle (6pt);
        \fill (18, 2) circle (6pt);
        \fill (7, -1) circle (6pt);
        \fill (7, 0) circle (6pt);
        \fill (7, 1) circle (6pt);
        \fill (7, 2) circle (6pt);
        \fill (8, 0) circle (6pt);
        \fill (8, 1) circle (6pt);
        \fill (8, 2) circle (6pt);
        \fill (9, 1) circle (6pt);
        \fill (9, 2) circle (6pt);
        \fill (10, 2) circle (6pt);
        \fill (-1, -1) circle (6pt);
        \fill (0, -1) circle (6pt);
        \fill (1, -1) circle (6pt);
        \fill (2, -1) circle (6pt);
        \fill (0, 0) circle (6pt);
        \fill (1, 0) circle (6pt);
        \fill (2, 0) circle (6pt);
        \fill (1, 1) circle (6pt);
        \fill (2, 1) circle (6pt);
        \fill (2, 2) circle (6pt);
	\end{pgfonlayer}
	\begin{pgfonlayer}{edgelayer}
		\draw (34.center) to (35.center);
		\draw (34.center) to (37.center);
		\draw (28.center) to (29.center);
		\draw (28.center) to (31.center);
		\draw (22.center) to (25.center);
		\draw (23.center) to (22.center);
		\draw (19.center) to (18.center);
		\draw (18.center) to (17.center);
		\draw [style=Arrow] (19.center) to (20.center);
		\draw [style=Arrow] (17.center) to (16.center);
		\draw [style=Arrow] (25.center) to (26.center);
		\draw [style=Arrow] (23.center) to (24.center);
		\draw [style=Arrow] (29.center) to (30.center);
		\draw [style=Arrow] (31.center) to (32.center);
		\draw [style=Arrow] (35.center) to (36.center);
		\draw [style=Arrow] (37.center) to (38.center);
	\end{pgfonlayer}
\end{tikzpicture}}
\caption{Sketch of the proof of the multiplication of harmonic sums (taken from \cite{Ablinger:2012ufz}).}
\label{fig:$S$-sumalg}
\end{figure}
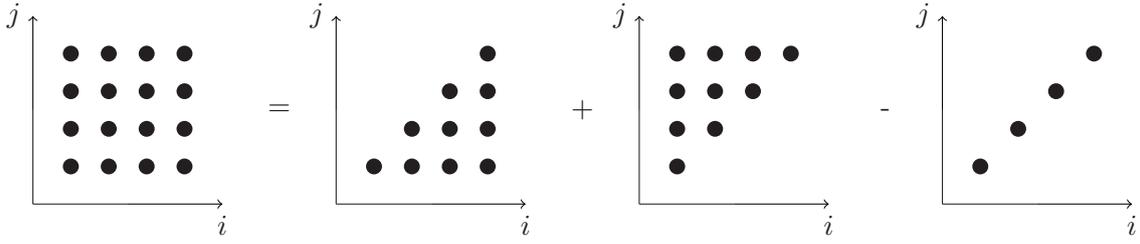
The proof of \eq{$S$-sum algebra} follows directly from the relation
\begin{equation}
    \sum_{i=1}^n \sum_{j=1}^n a_{ij} = \sum_{i=1}^n \sum_{j=1}^{i} a_{ij} + \sum_{j=1}^n \sum_{i=1}^{j} a_{ij} - \sum_{i=1}^n a_{ii} \,,
\end{equation}
which is sketched in fig.~\ref{fig:$S$-sumalg}. Eq.~\eqref{eq:$S$-sum algebra} has been implemented into an algorithm for the multiplication of two $S$-sums in \XSUMMER. \\

\noindent \textbf{Synchronization of arguments} \\
In order to apply \eq{$S$-sum algebra} one needs to have nested sums with the same argument. To synchronize the argument of the two $S$-sums one can recursively apply the following formula
\begin{align}\label{eq:Synchss}
    &S(n ; p_1, \dots, p_k; x_1, \dots, x_k)  \\
    &= S(n+1; p_1, \dots, p_k; x_1, \dots, x_k) - \frac{\text{sign}(p_1)^{n+1}x_1^{n+1}S(n+1; p_2, \dots, p_k; x_2, \dots, x_k)}{(n+1)^{\abs{p_1}}} \,. 
\nn\end{align}
Eq.~\eqref{eq:Synchss} has also been implemented in \XSUMMER. 

The only product that the combination of these two algorithms is not able to reduce further is
\begin{equation}
    S(i ; p_1, \dots, p_k; x_1, \dots, x_k) \cdot S(n-i ; p_1', \dots, p_k'; x_1', \dots, x_k') \,. \label{eq:Product $S$-sums}
\end{equation}
Consequently, in \IBIS, we have only solved inverse binomial sums that contain up to two  harmonic sums.\\

\noindent \textbf{Weights} \\
To perform checks on our results, we define a \textit{weight} $W$ and a \textit{product weight} $P$ of a generic $S$-sum
\begin{align}
    W(S(n;p_1,...,p_k;x_1,...,x_k)) &= \sum_{i=1}^k|p_i| \,, \\
    P(S(n;p_1,...,p_k;x_1,...,x_k)) &= \prod_{i=1}^k|x_i| \,.
\end{align}
Furthermore, it is useful to define the extended weight and extended product weight of the compound object of a sum and a denominator or monomial of which the exponent is equal to the argument of the sum:
\begin{equation}
W\biggl(a^n \cdot \frac{S(n;p_1,...,p_k;x_1,...,x_k)}{n^k}\bigg) = k+\sum_{i=1}^k|p_i| \,,
\end{equation}
\begin{equation}
P\biggl(a^n \cdot \frac{S(n;p_1,...,p_k;x_1,...,x_k)}{n^k}\bigg) = a\cdot \prod_{i=1}^k|x_i| \,.
\end{equation}
Similar definitions for the (extended) weight and product weight hold for $Z$-sums. \\

\noindent \textbf{Special cases} \\
$Z$-sums and $S$-sums generalize well-known mathematical objects. In particular, $S$-sums extend the concept of harmonic sums, which can appear within the inverse binomial sums that \IBIS is designed to solve. In the case of $x_1 = \dots = x_k = 1$, $S$-sums reduce to harmonic sums:
\begin{equation} \label{eq:harmonic}
    S(n; p_1, \dots, p_k; 1, \dots, 1) = S_{p_1, \dots, p_k}(n) \,,
\end{equation}
and $Z$-sums reduce to Euler-Zagier sums \cite{Euler:1776, Zagier:1994}:
\begin{equation}
    Z(n; p_1, \dots, p_k; 1, \dots, 1) = Z_{p_1, \dots, p_k}(n) \,.
\end{equation}
Below is an overview of some other important mathematical objects that are generalized by $S$-sums and $Z$-sums. For $n = \infty$, $Z$-sums reduce to the multiple polylogarithms of Goncharov \cite{goncharov2011multiplepolylogarithmscyclotomymodular}:
\begin{equation}
    Z(\infty; p_1, \dots, p_k; x_1, \dots, x_k) = \text{Li}_{p_k, \dots, p_1}(x_k, \dots, x_1) \,.
\end{equation}
The multiple polylogarithms of Goncharov include several special cases, such as classical polylogarithms $\text{Li}_n(x)$ \cite{Lewin:1981}, Nielsen's generalized polylogarithms \cite{Nielsen:1909}
\begin{equation}
    S_{n,p}(x) = \text{Li}_{1,\dots,1,n+1}(\underbrace{1,\dots,1}_{p-1},x) \,,
\end{equation}  
and the harmonic polylogarithms of Remiddi and Vermaseren \cite{Remiddi:1999ew}
\begin{equation}
    H_{p_1, \dots, p_k}(x) = \text{Li}_{p_k,\dots,p_1}(\underbrace{1,\dots,1}_{k-1},x)\,.
\end{equation}
For $n = \infty$ and $x_1 = \dots = x_k = 1$, the sum becomes a multiple zeta value \cite{borwein1999specialvaluesmultiplepolylogarithms}:
\begin{equation}
    Z(\infty; p_1, \dots, p_k; 1, \dots, 1) = \zeta(p_k, \dots, p_1) \,.
\end{equation}
These relationships demonstrate the broad applicability of $S$-sums and $Z$-sums in the context of modern mathematical and physical studies.

%%%%%%%%%%%%%%%%%%%%%%%%%%%%%%%%%%%%%%%%%%%%
\section{The recursion} \label{sec:Recursionsec}
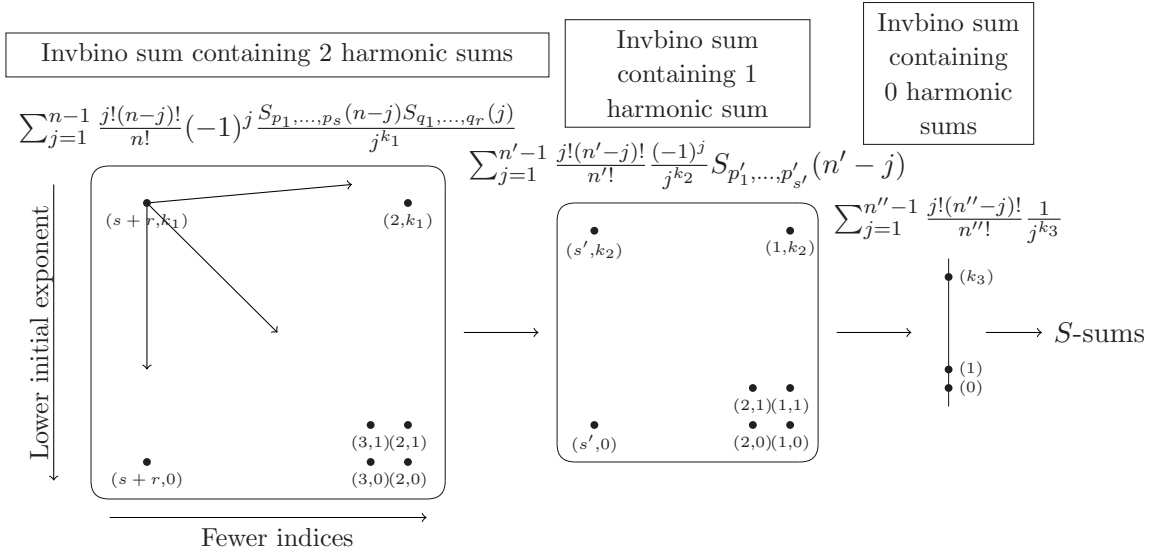
\begin{figure}[!t]
\centering
  \resizebox{\textwidth}{!}{\begin{tikzpicture}[scale=0.5]
	\begin{pgfonlayer}{nodelayer}
		\node [style=none] (0) at (-17.5, 4.5) {};
		\node [style=none] (1) at (-18, 4) {};
		\node [style=none] (2) at (-9, 4.5) {};
		\node [style=none] (3) at (-8.5, 4) {};
		\node [style=none] (4) at (-8.5, -4) {};
		\node [style=none] (5) at (-9, -4.5) {};
		\node [style=none] (6) at (-17.5, -4.5) {};
		\node [style=none] (7) at (-18, -4) {};
		\node [style=none] (8) at (-8, 0) {};
		\node [style=none] (9) at (-6, 0) {};
		\node [style=none] (10) at (-5.5, 3) {};
		\node [style=none] (11) at (-5, 3.5) {};
		\node [style=none] (12) at (-5.5, -3) {};
		\node [style=none] (13) at (-5, -3.5) {};
		\node [style=none] (14) at (1, -3.5) {};
		\node [style=none] (15) at (1.5, -3) {};
		\node [style=none] (16) at (1.5, 3) {};
		\node [style=none] (17) at (1, 3.5) {};
		\node [style=none] (18) at (2, 0) {};
		\node [style=none] (19) at (4, 0) {};
		\node [style=none] (20) at (5, 2) {};
		\node [style=none] (22) at (5, -2) {};
		\node [style=none] (28) at (6, 0) {};
		\node [style=none, label={right:$S$-sums}] (30) at (7.5, 0) {};
		\node [style=none, label={below:$\sum_{j=1}^{n-1} \frac{j!(n-j)!}{n!}(-1)^j\frac{S_{p_1,...,p_s}(n-j)S_{q_1,...,q_r}(j)}{j^{k_1}}$}] (37) at (-13.25, 6.5) {};
		\node [style=none, label={below:$\sum_{j=1}^{n'-1} \frac{j!(n'-j)!}{n'!}\frac{(-1)^j}{j^{k_2}}S_{p'_1,...,p'_{s'}}(n'-j)$}] (39) at (-2, 5.5) {};
		\node [style=none, label={below:$\sum_{j=1}^{n''-1} \frac{j!(n''-j)!}{n''!}\frac{1}{j^{k_3}}$}] (41) at (5, 4) {};
        \fill (-16.5, 3.5) circle (3pt);
        \fill (-9.5, 3.5) circle (3pt);
        \fill (-9.5, -3.5) circle (3pt);
        \fill (-9.5, -2.5) circle (3pt);
        \fill (-10.5, -2.5) circle (3pt);
        \fill (-10.5, -3.5) circle (3pt);
        \fill (-16.5, -3.5) circle (3pt);
        \fill (-4.5, 2.75) circle (3pt);
        \fill (0.75, 2.75) circle (3pt);
        \fill (0.75, -2.5) circle (3pt);
        \fill (-0.25, -1.5) circle (3pt);
        \fill (0.75, -1.5) circle (3pt);
        \fill (-0.25, -2.5) circle (3pt);
        \fill (-4.5, -2.5) circle (3pt);
        \fill (5, 1.5) circle (3pt);
        \fill (5, -1.5) circle (3pt);
        \fill (5, -1) circle (3pt);
		\node [font={\tiny}, anchor=north] (42) at (-16.5, 3.5) {($s+r$,$k_1$)};
		\node [font={\tiny}, anchor=north] (43) at (-9.5, 3.5) {($2$,$k_1$)};
		\node [font={\tiny}, anchor=north] (44) at (-9.5, -3.5) {($2$,$0$)};
		\node [font={\tiny}, anchor=north] (45) at (-9.5, -2.5) {($2$,$1$)};
		\node [font={\tiny}, anchor=north] (46) at (-10.5, -2.5) {($3$,$1$)};
		\node [font={\tiny}, anchor=north] (47) at (-10.5, -3.5) {($3$,$0$)};
		\node [font={\tiny}, anchor=north] (48) at (-16.5, -3.5) {($s+r$,$0$)};
		\node [font={\tiny}, anchor=north] (49) at (-4.5, 2.75) {($s'$,$k_2$)};
		\node [font={\tiny}, anchor=north] (50) at (0.75, 2.75) {($1$,$k_2$)};
		\node [font={\tiny}, anchor=north] (51) at (0.75, -2.5) {($1$,$0$)};
		\node [font={\tiny}, anchor=north] (52) at (-0.25, -1.5) {($2$,$1$)};
		\node [font={\tiny}, anchor=north] (53) at (0.75, -1.5) {($1$,$1$)};
		\node [font={\tiny}, anchor=north] (54) at (-0.25, -2.5) {($2$,$0$)};
		\node [font={\tiny}, anchor=north] (55) at (-4.5, -2.5) {($s'$,$0$)};
		\node [font={\tiny}, anchor=west] (56) at (5, 1.5) {($k_3$)};
		\node [font={\tiny}, anchor=west] (57) at (5, -1.5) {($0$)};
		\node [font={\tiny}, anchor=west] (58) at (5, -1) {($1$)};
		\node [style=none] (59) at (-17.5, -5) {};
		\node [below, font={\small}] (60) at (-13, -5) {Fewer indices};
		\node [style=none] (61) at (-9, -5) {};
		\node [style=none] (62) at (-19, -4) {};
		\node [style=none] (64) at (-19, 4) {};
		\node [style=none, rotate=90, font={\small}, anchor=south] (69) at (-19, 0) {Lower initial exponent};
		\node [style=none, draw, inner sep=4pt, font={\small}, text width=7cm, align=center] (70) at (-13, 7.5) {Invbino sum containing 2 harmonic sums};
		\node [style=none, font={\small}, draw, inner sep=4pt, text width=2cm, align=center] (71) at (5, 7) {Invbino sum containing 0 harmonic sums};
		\node [style=none, node font={\small}, draw, inner sep=4pt, text width=3cm, align=center] (72) at (-2, 7) {Invbino sum containing 1 harmonic sum};
        \node [style=none] (73) at (-16.5, -1) {};
        \node [style=none] (74) at (-13, 0) {};
        \node [style=none] (75) at (-11, 4) {};
        \node [style=none] (76) at (-16.5, 3.5) {};
	\end{pgfonlayer}
	\begin{pgfonlayer}{edgelayer}
		\draw (1.center) to (7.center);
		\draw (0.center) to (2.center);
		\draw (3.center) to (4.center);
		\draw (5.center) to (6.center);
		\draw [bend right=45, looseness=1.25] (7.center) to (6.center);
		\draw [bend left=45, looseness=1.25] (1.center) to (0.center);
		\draw [bend left=45, looseness=1.25] (2.center) to (3.center);
		\draw [bend right=45, looseness=1.25] (5.center) to (4.center);
		\draw [style=Arrow] (8.center) to (9.center);
		\draw (11.center) to (17.center);
		\draw (15.center) to (16.center);
		\draw (12.center) to (10.center);
		\draw (13.center) to (14.center);
		\draw [bend left=45, looseness=1.25] (17.center) to (16.center);
		\draw [bend left=45, looseness=1.25] (15.center) to (14.center);
		\draw [bend left=45, looseness=1.25] (13.center) to (12.center);
		\draw [bend left=45, looseness=1.25] (10.center) to (11.center);
		\draw [style=Arrow] (18.center) to (19.center);
		\draw (20.center) to (22.center);
		\draw [style=Arrow] (28.center) to (30.center);
		\draw [style=Arrow] (59.center) to (61.center);
		\draw [style=Arrow] (64.center) to (62.center);
        \draw [style=Arrow] (76.center) to (73.center);
        \draw [style=Arrow] (76.center) to (74.center);
        \draw [style=Arrow] (76.center) to (75.center);
	\end{pgfonlayer}
\end{tikzpicture}}
\caption{An illustration of how \IBIS reduces the complexity of an inverse binomial sum: We start with 2 harmonic sums (shown on the left). By recursion the complexity is reduced (direction of arrows), either lowering the initial exponent or the number of indices. Iterating, this reduces to inverse binomial sums with only a single harmonic sum (shown in the middle). Again recursions are used such that we end up with contributions only involving inverse binomial sums without harmonic sums, which are then expressed as $S$-sums (on the right). Note that the expressions are just illustrative, as \IBIS can handle many more classes of inverse binomial sums.
}
\label{fig:Invbino}
\end{figure}

%%%%%%%%%%%%%%%%%%%%%%%%%%%%%%%%%%%%%%%%%%%%
Our program applies recursion to progressively reduce the complexity of inverse binomial sums, ultimately resulting in a combination of $S$-sums. To illustrate what we mean by reducing complexity, consider a typical sign-altering inverse binomial sum that contains two harmonic sums:
\begin{equation}
    \sum_{j=1}^{n-1} \frac{j!(n-j)!}{n!}\,(-1)^j\,\frac{S_{p_1,...,p_s}(n-j)S_{q_1,...,q_r}(j)}{j^{k}} \,. \label{eq:example invbino}
\end{equation}
Here, we define $k$ as the \textit{initial exponent}, $n$ as the \textit{upper limit} and the $p_i, q_j$ as the \textit{indices}. The complexity of such an inverse binomial sum is reduced by expressing it in terms of other inverse binomial sums with either a lower initial exponent or fewer indices. For example, in the case of \eq{example invbino}, \IBIS first reduces the sum to inverse binomial sums containing one harmonic sum, then to sums with no harmonic sums, and ultimately to $S$-sums. This process is illustrated in fig.~\ref{fig:Invbino}.
It is important to note that in this recursion inverse binomial sums will sometimes be expressed in terms of sums with a higher initial exponent, but in these cases the resulting sums have fewer indices. However, an inverse binomial sum will never be expressed in terms of a sum with more indices. 

The rest of this section walks through each step of deriving the recursions, using a specific example. In \sec{telescoping} we set up the telescoping series that yields our recursion relation. This requires the synchronization of the different terms in the series, discussed in \sec{synch_tel}. For cases where the denominator is offset by a constant integer, we describe how to remove the offset in \sec{synch_denominator}. Finally, in \sec{validation} we discuss the various ways in which we have validated our results.
The following notation will be used here,
\begin{align} \label{eq:invbinofixeddef}
    G^{p_1,...,p_s}(k,n) &\equiv \sum_{j=1}^{n-1} \frac{j!(n-j)!}{n!}\frac{1}{j^k}S_{p_1,...,p_s}(n-j) \,, \nn \\ G_{q_1,...,q_r}(k,n) &\equiv \sum_{j=1}^{n-1} \frac{j!(n-j)!}{n!}\frac{1}{j^k}S_{q_1,...,q_r}(j) \,,
\end{align}
where the initial exponent and the upper limit are the arguments of the function $G$.

%~~~~~~~~~~~~~~~~~~~~~~~~~~~~~~~~~~~~~~~~~~~~~~~
\subsection{Telescoping} \label{sec:telescoping}
%~~~~~~~~~~~~~~~~~~~~~~~~~~~~~~~~~~~~~~~~~~~~~~~
The main concept that is used to derive the recursions is telescoping \cite{Schneider:2013uan}. 
Concretely, to find a recurrence relation for $G^{p_1,...,p_s}(k,n)$, we set up the telescoping series as follows:
\begin{equation}\label{eq:exampleTelescoping}
    G^{p_1,...,p_s}(k,n) = \sum_{i=1}^n \Bigl( G^{p_1,...,p_s}(k,i) - G^{p_1,...,p_s}(k,i-1) \Bigr) \,,
\end{equation}
where subsequent terms cancel in the sum on $i$, leaving only the left-hand side and $G^{p_1,...,p_s}(k,0)$.
Since $G^{p_1,...,p_s}(k,0)$ is an empty sum and thus equal to zero, the equivalence becomes evident. We set up this telescoping series with respect to $n$, because the whole point of our work is to solve for symbolic $n$ rather than restricting to specific numerical values. To proceed with this telescoping series, we need to derive an expression for $G^{p_1,...,p_s}(k,i) - G^{p_1,...,p_s}(k,i-1)$ where either the number of indices or the initial exponent of the inverse binomial sum has been reduced, which then yields a recursive relation that reduces the complexity of the inverse binomial sum.

To find an expression for $G^{p_1,...,p_s}(k,i) - G^{p_1,...,p_s}(k,i-1)$ of which all the terms have a lower complexity than $G^{p_1,...,p_s}(k,n)$, we can rewrite the inverse binomial factor as follows:
\begin{equation} \label{eq:inv_bin_rewrite}
    \frac{1}{\binom{i}{j}} = \frac{j!(i-j-1)!}{(i-1)!}\biggl( 1 - \frac{j}{i} \biggr) = \frac{1}{\binom{i-1}{j}} \bigg( 1 - \frac{j}{i} \bigg) \,.
\end{equation}
By doing so, we are able to rewrite the inverse binomial sum with $i$ as its upper limit into the same inverse binomial sum, but now with $i-1$ as its upper limit. For our example in \eq{exampleTelescoping}, this yields
\begin{align}
    G^{p_1,...,p_s}(k,i)
    &=\sum_{j=1}^{i-1} \frac{1}{\binom{i}{j}}\, \frac{S_{p_1,...,p_s}(i-j)}{j^k} \nn \\
    &= \sum_{j=1}^{i-1} \frac{1}{\binom{i-1}{j}} \bigg( 1-\frac{j}{i} \bigg) \frac{1}{j^k} S_{p_1,...,p_s}(i-j) \,.\label{eq:exampleTelescoping2}
\end{align}
However, this does not yet give a proper recursion as the arguments of $S_{p_1,...,p_s}(i-j)$ and $\frac{1}{\binom{i-1}{j}}$ are not aligned: $S_{p_1,...,p_s}(i-j)$ contains $i$, while the binomial term contains $i-1$. As a result, we can't directly express the right-hand side in terms of the $G$ functions. We will discuss in \sec{synch_tel} how to remedy this by using \textit{synchronization}. 

We now comment on the reason for choosing to work with $S$-sums instead of $Z$-sums. The upper bound $n$ of the sum on $i$ in \eq{exampleTelescoping} is equal to the argument $n$ on the left-hand side, which naturally results in $S$-sums. A recursion that would result in $Z$-sums would use the following telescoping series instead:
\begin{equation}
    G^{p_1,...,p_s}(k,n) = \sum_{i=1}^{n-1} \Bigl( G^{p_1,...,p_s}(k,i+1) - G^{p_1,...,p_s}(k,i) \Bigr) \,,
\end{equation}
However, we found early on by studying some examples that using $S$-sums yielded shorter expressions for the recursions than with $Z$-sums.

%~~~~~~~~~~~~~~~~~~~~~~~~~~~~~~~~~~~~~~~~~~~~
\subsection{Synchronization of the telescope}
\label{sec:synch_tel}
%~~~~~~~~~~~~~~~~~~~~~~~~~~~~~~~~~~~~~~~~~~~~
When the summand contains multiple terms, a common challenge is that their arguments may differ in such a way that we are not able to directly write the sum in terms of expressions that we know (i.e.~the $G$ functions). Here we discuss how to address this using synchronization, illustrating this first for a toy example, then a realistic example used in \IBIS, followed by a discussion of the subtleties that appear in other cases.\\

\noindent \textbf{Toy example}\\
We start by looking at a simple example to show how a slight change in one of the terms in the summands can mean that the nested sums cannot be integrated in the $S$-sum framework:
\begin{align} \label{eq:toy_example}
    \sum_{i_1=1}^n &\frac{(\text{sign}(p_1))^{i_1}x_1^{i_1}}{i_1^{|p_1|}} \sum_{i_2=1}^{i_1} \frac{(\text{sign}(p_2))^{i_2+1}x_2^{i_2-2}}{i_2^{|p_2|}}  
    \sum_{i_3=1}^{i_2} \frac{(\text{sign}(p_3))^{i_3}x_3^{i_3}}{i_3^{|p_3|}} S(i_3;p_4,p_5;x_4,x_5) \nn \\
    &= \sum_{i_1=1}^n \frac{(\text{sign}(p_1))^{i_1}x_1^{i_1}}{i_1^{|p_1|}} \sum_{i_2=1}^{i_1} \frac{(\text{sign}(p_2))^{i_2+1}x_2^{i_2-2}}{i_2^{|p_2|}} S(i_2;p_3,p_4,p_5;x_3,x_4,x_5) \,.
\end{align}
Due to the exponents of $\text{sign}(p_2)$ and $x_2$ in the numerator of the second sum not matching up with the summation variable $i_2$, only the third sum (over $i_3$) can directly be expressed as an $S$-sum. In the current form, the rest of the sums (over $i_2$ and $i_1$) would have to be performed explicitly by \FORM using \SUMMER. To avoid this, it is important to adjust the arguments of the objects in the summand so that they match, which is called synchronization. 
In this example this can simply be achieved by writing \eq{toy_example} as
\begin{align}
    \frac{\text{sign}(p_2)}{x_2^2}\sum_{i_1=1}^n \frac{(\text{sign}(p_1))^{i_1}x_1^{i_1}}{i_1^{|p_1|}} &\sum_{i_2=1}^{i_1} \frac{(\text{sign}(p_2))^{i_2}x_2^{i_2}}{i_2^{|p_2|}} S(i_2;p_3,p_4,p_5;x_3,x_4,x_5) \nn \\
    &= \frac{\text{sign}(p_2)}{x_2^2} S(n;p_1,p_2,p_3,p_4,p_5;x_1,x_2,x_3,x_4,x_5) \,.
\end{align} \\

\noindent \textbf{Realistic example} \\
To see how we apply this concept of synchronization to find the recursions for \IBIS, let us show how the terms for our example in \eq{exampleTelescoping2} can be synchronized. Suppose that $k>1$ and $p_1 \geq 0$, we get
\begin{align} \label{eq:G_tele}
    G^{p_1,...,p_s}(k,i)
    &= \sum_{j=1}^{i-1} \frac{1}{\binom{i-1}{j}} \frac{1}{j^k} \bigg( S_{p_1,...,p_s}(i-j-1) + \frac{S_{p_2,...,p_s}(i-j)}{(i-j)^{p_1}} \bigg) \\
    &\hspace{1cm}- \sum_{j=1}^{i-1} \frac{j!(i-j-1)!}{i!} \frac{S_{p_1,...,p_s}(i-j)}{j^{k-1}} \nn \\
%    &= \sum_{j=1}^{n-1} \frac{1}{\binom{n-1}{j}} \frac{1}{j^k} \bigg( S_{p_1,...,p_s}(n-j-1) + \frac{S_{p_2,...,p_s}(n-j)}{(n-j)^{p_1}} \cdot \frac{n(n-j)}{n(n-j)} \bigg) \nn\\
%    &\hspace{1cm}- \sum_{j=1}^{n-1} \frac{j!(n-j-1)!}{n!} \frac{S_{p_1,...,p_s}(n-j)}{j^{k-1}} \cdot \frac{n-j}{n-j} \nn \\
    &= G^{p_1,...,p_s}(k,i-1) + \frac{(i-1)!0!}{(i-1)!} \frac{1}{(i-1)^k} S_{p_1,...,p_s}(0) \nn\\
    &\hspace{1cm}+ i\sum_{j=1}^{i-1} \frac{j!(i-j)!}{i!} \frac{1}{j^k} \frac{S_{p_2,...,p_s}(i-j)}{(i-j)^{p_1+1}}  - \sum_{j=1}^{i-1} \frac{j!(i-j)!}{i!} \frac{S_{p_1,...,p_s}(i-j)}{j^{k-1}(i-j)} \,, \nn
\end{align}
where we used \eq{Synchss} in the first step of this derivation and that $S_{p_1,...,p_s}(0)=0$ for all $p_1,...,p_s \in \mathbb{Z}$. As shown, we have successfully synchronized the arguments of the binomial and $S$-sum, allowing us to identify $G^{p_1,...,p_s}(k,i-1)$ along with other terms that still require synchronization. 

Since we restrict inverse binomial sums to denominators of the form $\frac{1}{j^k}$ rather than $\frac{1}{j^k(i-j)^{k'}}$, the next step is to rewrite the remaining terms in \eq{G_tele} in terms of $G$ functions. Using the following general expression to split the fractions,
\begin{align}
    \frac{1}{(i+x)^m(i+y)^k}
    &= \sum_{a=1}^m\binom{m+k-1-a}{k-1}\frac{(-1)^{m-a}}{(y-x)^{m+k-a}(i+x)^a} \nn \\
    &\hspace{1cm}+ \sum_{a=1}^k\binom{m+k-1-a}{m-1}\frac{(-1)^{m}}{(y-x)^{m+k-a}(i+y)^a} \,,\label{eq:splitFractions}
\end{align}
we obtain
\begin{align}
    &G^{p_1,...,p_s}(k,i)    \nn \\
%    &= G^{p_1,...,p_s}(k,n-1) + n\sum_{j=1}^{n-1} \frac{j!(n-j)!}{n!} S_{p_2,...,p_s}(n-j) \cdot \nn\\
%    &\hspace{1cm}\bigg( \sum_{a=1}^k\binom{k+p_1+1-1-a}{p_1+1-1}\frac{1}{n^{k+p_1+1-a}j^a}
%    \nn\\
%    &\hspace{1cm}+ \sum_{a=1}^{p_1+1}\binom{k+p_1+1-1-a}{k-1}\frac{1}{n^{k+p_1+1-a}(n-j)^a} \bigg)  \nn\\
%    &\hspace{1cm}- \sum_{j=1}^{n-1} \frac{j!(n-j)!}{n!} S_{p_1,...,p_s}(n-j) \bigg( \sum_{a=1}^{k-1}\frac{1}{n^{k-1+1-a}j^a}
%    + \frac{1}{n^{k-1}(n-j)} \bigg)\nn \\
    &= G^{p_1,...,p_s}(k,i-1) + \sum_{a=1}^k\binom{k+p_1-a}{p_1} \frac{1}{i^{k+p_1-a}}\sum_{j=1}^{i-1} \frac{j!(i-j)!}{i!}  \frac{S_{p_2,...,p_s}(i-j)}{j^a}
    \nn\\
    &\hspace{1cm}+ \sum_{a=1}^{p_1+1}\binom{k+p_1-a}{k-1} \frac{1}{i^{k+p_1-a}} \sum_{j=1}^{i-1} \frac{j!(i-j)!}{i!}  \frac{S_{p_2,...,p_s}(i-j)}{(i-j)^a}  \nn\\
    &\hspace{1cm}- \sum_{a=1}^{k-1} \frac{1}{i^{k-a}} \sum_{j=1}^{i-1} \frac{j!(i-j)!}{i!} \frac{S_{p_1,...,p_s}(i-j)}{j^a}
    - \frac{1}{i^{k-1}} \sum_{j=1}^{i-1} \frac{j!(i-j)!}{i!} \frac{S_{p_1,...,p_s}(i-j)}{(i-j)} \nn \\
    &= G^{p_1,...,p_s}(k,i-1) + \sum_{a=1}^k\binom{k+p_1-a}{p_1} \frac{1}{i^{k+p_1-a}}G^{p_2,...,p_s}(a,i) - \frac{1}{i^{k-1}} G_{p_1,...,p_s}(1,i) 
    \nn\\
    &\hspace{1cm}+ \sum_{a=1}^{p_1+1}\binom{k+p_1-a}{k-1} \frac{1}{i^{k+p_1-a}} G_{p_2,...,p_s}(a,i) - \sum_{a=1}^{k-1} \frac{1}{i^{k-a}} G^{p_1,...,p_s}(a,i) \,.\label{eq:Syncinvbino}
\end{align}
In the last line, we performed a change of the summation variable, substituting $j \to i-j$ for the third and fifth term on the right-hand side. This transformation caused the indices of the inverse binomial sums to shift from lower to upper indices. By synchronizing all elements in the summands, we rewrote the expressions in terms of recognizable components (i.e. inverse binomial sums like in \eq{invbinofixeddef}). This approach minimizes the number of sums that need to be evaluated. 

The final step involves setting up the telescoping series:
\begin{align}
    G^{p_1,...,p_s}(k,n) &= \sum_{i=1}^n \bigg( G^{p_1,...,p_s}(k,i) - G^{p_1,...,p_s}(k,i-1) \bigg) \nn \\
%    &= \sum_{i=2}^n \bigg( \sum_{a=1}^k\binom{k+p_1-a}{p_1} \frac{1}{i^{k+p_1-a}}G^{p_2,...,p_s}(a,i)+ \sum_{a=1}^{p_1+1}\binom{k+p_1-a}{k-1} \frac{1}{i^{k+p_1-a}} G_{p_2,...,p_s}(a,i) \nn\\
%    &\hspace{1cm}- \sum_{a=1}^{k-1} \frac{1}{i^{k-a}} G^{p_1,...,p_s}(a,i)
%    - \frac{1}{i^{k-1}} G_{p_1,...,p_s}(1,i) \bigg) \nn \\
    &= \sum_{a=1}^k\binom{k+p_1-a}{p_1} \sum_{i=1}^n\frac{1}{i^{k+p_1-a}}G^{p_2,...,p_s}(a,i) - \sum_{a=1}^{k-1} \sum_{i=1}^n\frac{1}{i^{k-a}} G^{p_1,...,p_s}(a,i) \nn\\
    &\hspace{0.8cm}+ \sum_{a=1}^{p_1+1}\binom{k+p_1-a}{k-1} \sum_{i=1}^n\frac{1}{i^{k+p_1-a}} G_{p_2,...,p_s}(a,i)
    - \sum_{i=1}^n\frac{1}{i^{k-1}} G_{p_1,...,p_s}(1,i) \,.\label{eq:finalrec}
\end{align}
By establishing this telescoping series, we have derived a recursion expressed in terms of recognizable components with either upper or lower indices. The inverse binomial sums on the right-hand side have a lower complexity, meaning that they have either fewer indices or a reduced initial exponent, guaranteeing that the recursion will terminate. \\

\noindent \textbf{Subtleties} \\
We have carried out this process of setting up the telescoping series after finding an expression for the summand for every different type of inverse binomial sum listed in tab.~\ref{table:possible sums}. The following subtleties are worth mentioning: First of all, during the synchronization process of these  recursions, we encountered terms where the upper limit of an inverse binomial sum was not synchronized with its denominator. To synchronize these two elements, a different technique was required, which
is illustrated here for the following term:
\begin{align}
    \frac{n}{2^{n+1}} \sum_{i=1}^{n} \frac{2^{i}}{i}G_{p_2,...,p_s}(p_1,i-1) \,.\label{eq:interestingterm}
\end{align}
To synchronize the inverse binomial sum with the denominator, we need to find an expression for $G_{p_2,...,p_s}(p_1,i-1)$ in terms of inverse binomial sums with $i$ as the upper limit. While constructing the telescoping series, we derived an expression for $G^{p_1,...,p_s}(k,i) - G^{p_1,...,p_s}(k,i-1)$ in terms of inverse binomial sums with $i$ as the upper limit. By rewriting this expression, we can express $G^{p_1,...,p_s}(k,i-1)$ in terms of inverse binomial sums with $i$ as its upper limit, as desired. Thus, let us rewrite \eq{Syncinvbino} as follows:
\begin{align} 
    &G^{p_1,...,p_s}(k,i-1) \nn \\
    &= G^{p_1,...,p_s}(k,i) - \sum_{a=1}^k\binom{k+p_1-a}{p_1} \frac{1}{i^{k+p_1-a}}G^{p_2,...,p_s}(a,i) + \frac{1}{i^{k-1}} G_{p_1,...,p_s}(1,i)
    \nn\\
    &\hspace{1cm}- \sum_{a=1}^{p_1+1}\binom{k+p_1-a}{k-1} \frac{1}{i^{k+p_1-a}} G_{p_2,...,p_s}(a,i) + \sum_{a=1}^{k-1} \frac{1}{i^{k-a}} G^{p_1,...,p_s}(a,i) \,. \label{eq:syncinvbinofinal}
\end{align}
This relation for inverse binomial sums is similar to the relation between $S$-sums with shifted arguments in \eq{Synchss}. To fully synchronize the inverse binomial sum with the denominator, all that is left is to substitute \eq{syncinvbinofinal} into \eq{interestingterm}. For the case of $p_1>1$ and $p_2\geq0$ we therefore find that
\begin{align}
    &\frac{n}{2^{n+1}} \sum_{i=1}^{n} \frac{2^{i}}{i}G_{p_2,...,p_s}(p_1,i-1) \nn \\
%    &= \frac{n}{2^{n+1}} \sum_{i=1}^{n} \frac{2^{i}}{i} \bigg( G^{p_2,...,p_s}(p_1,i) - \sum_{a=1}^{p_1}\binom{p_1+p_2-a}{p_2} \frac{1}{i^{p_1+p_2-a}}G^{p_3,...,p_s}(a,i) \nn\\
%    &\hspace{1cm}- \sum_{a=1}^{p_2+1}\binom{p_1+p_2-a}{p_1-1} \frac{1}{i^{p_1+p_2-a}} G_{p_3,...,p_s}(a,i)\nn\\
%    &\hspace{1cm}+ \sum_{a=1}^{p_1-1} \frac{1}{i^{p_1-a}} G^{p_2,...,p_s}(a,i)
%    + \frac{1}{i^{p_1-1}} G_{p_2,...,p_s}(1,i) \bigg) \nn \\
    &= \frac{n}{2^{n+1}} \sum_{i=1}^{n} \frac{2^{i}}{i} G^{p_2,...,p_s}(p_1,i) - \sum_{a=1}^{p_1}\binom{p_1+p_2-a}{p_2} \frac{n}{2^{n+1}} \sum_{i=1}^{n} \frac{2^{i}}{i^{p_1+p_2+1-a}}G^{p_3,...,p_s}(a,i)
    \nn\\
    &\hspace{1cm}- \sum_{a=1}^{p_2+1}\binom{p_1+p_2-a}{p_1-1}\frac{n}{2^{n+1}} \sum_{i=1}^{n} \frac{2^{i}}{i^{p_1+p_2+1-a}} G_{p_3,...,p_s}(a,i) \nn\\
    &\hspace{1cm}+ \sum_{a=1}^{p_1-1} \frac{n}{2^{n+1}} \sum_{i=1}^{n} \frac{2^{i}}{i^{p_1+1-a}} G^{p_2,...,p_s}(a,i) + \frac{n}{2^{n+1}} \sum_{i=1}^{n} \frac{2^{i}}{i^{p_1}} G_{p_2,...,p_s}(1,i) \,. \label{eq:inttermfinal}
\end{align}
As one can see, the upper limits in the argument of the $G$ functions representing the inverse binomial sums on the right-hand side are now synchronized with their corresponding denominators. 

A second interesting subtlety is that we had to distinguish numerous cases. For example for \eq{inttermfinal}, the reader can see that specifying $p_1>1$ and $p_2 \geq 0$ was required due to the appearance of the sign function in the definition of $S$-sums in \eq{$S$-sumdef}. In the case of inverse binomial sums containing two harmonic sums, this process ultimately required distinguishing 124 distinct cases, each with unique conditions on the indices $p_i$, and implementing them systematically in \FORM. This intensive case-by-case approach was essential for handling the diversity of possible inputs. For interested readers, we have all computations for these recursions available on request.

%~~~~~~~~~~~~~~~~~~~~~~~~~~~~~~~~~~~~~~~~~~~~~~
\subsection{Synchronization of the denominator}
\label{sec:synch_denominator}
%~~~~~~~~~~~~~~~~~~~~~~~~~~~~~~~~~~~~~~~~~~~~~~
Our package includes a feature that handles inverse binomial sums where the denominator is not synchronized with the rest of the summand, i.e. sums like
\begin{equation}
    G^{p_1,...,p_s}(k,n)_c 
    \equiv \sum_{j=1}^{n-1} \frac{j!(n-j)!}{n!}\frac{1}{(j+c)^k}S_{p_1,...,p_s}(n-j) \,,
\end{equation}
with $c$ a positive integer\footnote{In the manual cases are discussed where the lower bound on the sum on $j$ is larger than 1, allowing other values of $c$ subject to $j+c>0$.}. These shifted denominators can for example arise when rewriting the arguments of the binomial by using identities like \eq{inv_bin_rewrite}.

To address these types of inverse binomial sums, we iteratively reduce the value of $c$ to zero. This reduction is achieved through recursions obtained by changing the summation variable:
\begin{equation}
    G^{p_1,...,p_s}(k,n)_c = \sum_{j=2}^{n} \frac{(j-1)!(n+1-j)!}{n!}\frac{1}{(j+c-1)^k}S_{p_1,...,p_s}(n+1-j) \cdot \frac{j}{j} \,.
\end{equation}
After this step, the remaining elements only need to be synchronized and there is no need to set up a telescoping series. We therefore focus solely on formulating a recursion for $G^{p_1,...,p_s}(k,n)_c$ that can iteratively reduce the value of $c$ to zero, and is expressed in terms of recognizable components. This process is carried out as follows:
\begin{align}
    G^{p_1,...,p_s}(k,n)_c &= (n+1)\sum_{j=1}^{n} \frac{j!(n+1-j)!}{(n+1)!}\frac{1}{j(j+c-1)^k}S_{p_1,...,p_s}(n+1-j) \nn\\
    &\hspace{1cm}- (n+1)\frac{1!(n+1-1)!}{(n+1)!}\frac{1}{1(1+c-1)^k}S_{p_1,...,p_s}(n+1-1) \,,
\end{align}
after which we can split the fractions
\begin{align} \label{eq:ctoc-1final}
    &G^{p_1,...,p_s}(k,n)_c \\
    &= (n+1)\sum_{j=1}^{n} \frac{j!(n+1-j)!}{(n+1)!}S_{p_1,...,p_s}(n+1-j) \bigg( \frac{1}{(c-1)^{k}j} \nn \\
    &\hspace{1cm}- \sum_{a=1}^k \frac{1}{(c-1)^{k+1-a}(j+c-1)^{a}} \bigg) -\frac{S_{p_1,...,p_s}(n)}{c^k} \nn\\
%    &= \frac{(n+1)}{(c-1)^k}\sum_{j=1}^{n} \frac{j!(n+1-j)!}{(n+1)!}\frac{1}{j}S_{p_1,...,p_s}(n+1-j) -\frac{S_{p_1,...,p_s}(n)}{c^k}\nn\\
    &\hspace{1cm}- \sum_{a=1}^k\frac{n+1}{(c-1)^{k+1-a}}\sum_{j=1}^{n} \frac{j!(n+1-j)!}{(n+1)!}\frac{1}{(j+c-1)^{a}}S_{p_1,...,p_s}(n+1-j) \nn\\
    &= \frac{(n+1)}{(c-1)^k}G^{p_1,...,p_s}(1,n+1)_0 - \sum_{a=1}^k\frac{n+1}{(c-1)^{k+1-a}}G^{p_1,...,p_s}(a,n+1)_{c-1} -\frac{S_{p_1,...,p_s}(n)}{c^k} \,. 
\nn\end{align}
As one can see, we have successfully derived a recursion that rewrites the original inverse binomial sum into sums with a lower value of $c$. By applying this process iteratively, we can rewrite the original inverse binomial sums into sums where the denominator and the arguments of the harmonic sums are synchronized.

%~~~~~~~~~~~~~~~~~~~~~~
\subsection{Validation}
\label{sec:validation}
%~~~~~~~~~~~~~~~~~~~~~~
To ensure that the derived recursion terminates correctly, it is important to verify that all inverse binomial sums in the recursion have either fewer indices than the original sum or a lower initial exponent. In our example we saw that this condition was satisfied. 

Finally, to check for any potential errors in the construction of the algorithms we have performed three additional checks:
\begin{enumerate}
    \item For every recursion, the extended weight of the terms on the left-hand side must be equal to the extended weight of the terms on the right-hand side. In our example this means that the extended weight of the terms on the right-hand side of \eq{finalrec} must all be equal to
    \begin{equation}
        W(G^{p_1,...,p_s}(k,n)) = k + \sum_{i=1}^s \abs{p_i} \,,
    \end{equation}
    which the reader can verify.
    \item For every recursion, the extended product weight of all the terms on the left-hand side must equal the extended product weight of all the terms on the right-hand side. Since we only allow the inverse binomial sums to contain harmonic sums, the extended product weight of the inverse binomial sums that we solve is always equal to 1 (i.e. $P(G^{p_1,...,p_s}(k,n)) = 1$). This implies that all terms on the right-hand side of the recursion must also have an extended product weight of 1, which can be easily verified for \eq{finalrec}.
    \item The strongest check we used to verify the correctness of the recursions implemented in \IBIS involved selecting a representative inverse binomial sum for each type listed in tab.~\ref{table:possible sums} and for each special case (including the 124 cases for two $S$-sums with $k=0$ mentioned in \sec{synch_tel}). By assigning numerical values to their upper limits, we obtained an exact answer in the form of fractions from \IBIS. These fractions were then compared with the fractions obtained from a direct evaluation of the original sum. Their equality confirmed the validity of all recursions included in \IBIS.
\end{enumerate}

%%%%%%%%%%%%%%%%%%%%%%%%%%
\section{\IBIS usage}
\label{sec:usage}
%%%%%%%%%%%%%%%%%%%%%%%%%%

With \IBIS, users can solve inverse binomial sums in terms of $S$-sums. The inverse binomial sums are of the form $\sum_{j=1}^{n-1}\, 1/\binom{n}{j} \cdots$,
where we specify the supported additional terms ($\cdots$) in tab.~\ref{table:possible sums}, classified according to their sign behavior and the number of harmonic sums they contain. In sec.~\ref{sec:inputOutput} we illustrate how to provide \IBIS with input using a few examples,  discuss some relevant \FORM notation used by \SUMMER and \XSUMMER, and the output \IBIS produces. In sec.~\ref{sec:examples}, we show how to perform the sums in the introduction using \IBIS. Finally, in sec.~\ref{sec:speed} we benchmark the speed of \IBIS. \\

\subsection{Input and output}
\label{sec:inputOutput}
\noindent \textbf{Notation} \\
\SUMMER can deal with harmonic sums and Euler-Zagier sums and their \FORM notation is
\begin{align}
S_{p_1,...,p_k}(n) \quad &\to \quad \texttt{S(R(p1,...,pk),n)} \nn \\
Z_{p_1,...,p_k}(n)\quad  &\to \quad \texttt{Z(R(p1,...,pk),n)}
\end{align}
\XSUMMER can handle $S$-sums and $Z$-sums with arbitrary $x$ dependence. Their \FORM notation is
\begin{align}
S(n;p_1,...,p_k;x_1,...,x_k)\quad &\to\quad\texttt{S(R(p1,...,pk),X(x1,...,xk),n)} \nn \\
Z(n;p_1,...,p_k;x_1,...,x_k)\quad &\to\quad\texttt{Z(R(p1,...,pk),X(x1,...,xk),n)}
\end{align}
For detailed instructions on the specific \SUMMER and \XSUMMER procedures available for preparing inverse binomial sums for \IBIS, we refer the reader to our manual. \\

\noindent \textbf{Input} \\
We now specify how the user should input the inverse binomial sum, by providing a few examples. First, we consider the following inverse binomial sum containing no harmonic sum:
\begin{equation}
    \sum_{j=1}^{n-1} \frac{j!(n-j)!}{n!}\frac{1}{(j+4)^3} \,.
\end{equation}
To express this sum in terms of $S$-sums using \IBIS, the user needs to input it in \FORM as follows:
\begin{align}
    \texttt{sum(j,1,n-1)*invbino(n,j)*den(j+4)\textasciicircum 3} 
\end{align}
Next, consider solving an inverse binomial sum that includes one harmonic sum:
\begin{equation}
    \sum_{j=1}^{n-1} \frac{j!(n-j)!}{n!}\frac{1}{j}S_{2,2,3}(n-j)
\,.\end{equation}
For this case, the \FORM input should look as follows:
\begin{align}
    \texttt{sum(j,1,n-1)*invbino(n,j)*den(j)*S(R(2,2,3),n-j)}
\end{align}
Finally, suppose we want to solve an inverse binomial sum containing two harmonic sums:
\begin{equation}
    \sum_{j=3}^{n-1} \frac{j!(n-j)!}{n!}\frac{(-1)^j}{(j-2)^4}S_{2,4,5}(n-j)S_{1,3}(j) \,.
\end{equation}
The input in \FORM should then be written as:
\begin{align}
    \texttt{sum(j,3,n-1)*invbino(n,j)}&\texttt{*den(j-2)\textasciicircum 4*sign(j)*} \nn \\
    &\texttt{S(R(2,4,5),n-j)*S(R(1,3),j)}
\end{align}

For additional examples of how to format inverse binomial sums and utilize \IBIS within your \FORM code, we refer the reader to the \IBIS manual, which is available in the GitHub repository alongside the \IBIS code. The manual also includes an extended list of inverse binomial sums that can be solved using \IBIS beyond those specified in tab.~\ref{table:possible sums}. This includes inverse binomial sums containing more than two harmonic sums, inverse binomial sums containing harmonic sums with unsynchronized arguments and a lot of other variations. So if your inverse binomial sum differs from the ones listed in tab.~\ref{table:possible sums}, it is still possible that \IBIS is able to solve it. \\

\begin{figure}[!t]
\centering
\resizebox{\textwidth}{!}{\begin{tikzpicture}[scale=0.5]
	\begin{pgfonlayer}{nodelayer}
		\node [style=none, font={\small}, draw, inner sep=4pt, text width=2cm, align=center] (71) at (-11, 1.75) {Invbino sum};
		\node [style=none] (0) at (-6.25, 2) {};
		\node [style=none] (1) at (-5.75, 2.5) {};
		\node [style=none] (2) at (-6.25, 1.5) {};
		\node [style=none] (3) at (-5.75, 1) {};
		\node [style=none] (4) at (-3.25, 1) {};
		\node [style=none] (5) at (-2.75, 1.5) {};
		\node [style=none] (6) at (-2.75, 2) {};
		\node [style=none] (7) at (-3.25, 2.5) {};
		\node [style=none] (10) at (-2.25, 1.75) {};
		\node [style=none] (12) at (-0.75, 1.75) {};
		\node [style=none] (13) at (-6.75, 1.75) {};
		\node [style=none] (14) at (-8.25, 1.75) {};
		\node [style=none] (15) at (-6, 0.5) {};
		\node [style=none] (16) at (-6, -0.5) {};
		\node [style=none] (17) at (-4.5, 1.75) {Interface};
		\node [style=none] (18) at (-0.25, 2) {};
		\node [style=none] (19) at (-0.25, 1.5) {};
		\node [style=none] (20) at (0.25, 1) {};
		\node [style=none] (21) at (0.25, 2.5) {};
		\node [style=none] (22) at (2.25, 2.5) {};
		\node [style=none] (23) at (2.75, 2) {};
		\node [style=none] (24) at (2.75, 1.5) {};
		\node [style=none] (25) at (2.25, 1) {};
		\node [style=none] (26) at (3.25, 1.75) {};
		\node [style=none] (27) at (7.25, 1.75) {};
		\node [style=none] (28) at (3.25, 0.25) {};
		\node [style=none] (29) at (-5.5, -1) {};
		\node [style=none] (30) at (-5, -0.75) {};
		\node [style=none] (31) at (-5, -1.25) {};
		\node [style=none] (32) at (-4.5, -1.75) {};
		\node [style=none] (33) at (-4.5, -0.25) {};
		\node [style=none] (39) at (2, 0.5) {};
		\node [style=none] (40) at (2, -0.5) {};
		\node [style=none] (41) at (1.5, -1) {};
		\node [style=none] (42) at (1, -0.75) {};
		\node [style=none] (43) at (1, -1.25) {};
		\node [style=none] (44) at (0.5, -0.25) {};
		\node [style=none] (45) at (0.5, -1.75) {};
		\node [style=none] (46) at (1.25, 1.75) {Invbino};
		\node [style=none] (47) at (-2, -1) {Synchronization};
		\node [style=none] (72) at (3.75, -0.25) {};
		\node [style=none, text width=1cm, label={above:symbolic n}] (73) at (5.25, 1.75) {};
		\node [style=none] (74) at (3.75, -0.25) {};
		\node [style=none, font={\small}, draw, inner sep=4pt, text width=1.25cm, align=center] (76) at (9.25, 1.75) {$S$-sums};
		\node [style=none, font={\small}, draw, inner sep=4pt, text width=3.75cm, align=center] (78) at (11.75, -0.25) {SubsS (XSUMMER)};
		\node [style=none] (79) at (16, -0.25) {};
		\node [style=none] (80) at (17, -0.25) {};
		\node [style=none, font={\small}, draw, inner sep=4pt, text width=1.5cm, align=center] (81) at (19, -0.25) {numerical result};
		\node [style=none, text width=0.5cm, label={above:$c=0$}] (82) at (-1.5, 1.75) {};
		\node [style=none, text width=1cm, label={below:$c>0$}] (83) at (-6, -1) {};
		\node [style=none, text width=1cm, label={below:$c=0$}] (84) at (2, -1) {};
		\node [style=none] (86) at (7.25, -0.25) {};
		\node [style=none, text width=1cm, label={below:numerical n}] (87) at (5.25, -0.25) {};
	\end{pgfonlayer}
	\begin{pgfonlayer}{edgelayer}
		\draw (1.center) to (7.center);
		\draw [bend left=45, looseness=1.25] (7.center) to (6.center);
		\draw (6.center) to (5.center);
		\draw [bend left=45, looseness=1.25] (5.center) to (4.center);
		\draw (4.center) to (3.center);
		\draw [bend right=315, looseness=1.25] (3.center) to (2.center);
		\draw (2.center) to (0.center);
		\draw [bend left=45, looseness=1.25] (0.center) to (1.center);
		\draw (30.center) to (31.center);
		\draw [bend right=45, looseness=1.25] (31.center) to (32.center);
		\draw (32.center) to (45.center);
		\draw [bend right=45, looseness=1.25] (45.center) to (43.center);
		\draw (43.center) to (42.center);
		\draw [bend right=45, looseness=1.25] (42.center) to (44.center);
		\draw (44.center) to (33.center);
		\draw [bend right=45, looseness=1.25] (33.center) to (30.center);
		\draw [bend left=45, looseness=1.25] (18.center) to (21.center);
		\draw (21.center) to (22.center);
		\draw [bend left=45, looseness=1.25] (22.center) to (23.center);
		\draw (23.center) to (24.center);
		\draw [bend left=45, looseness=1.25] (24.center) to (25.center);
		\draw (25.center) to (20.center);
		\draw [bend left=45, looseness=1.25] (20.center) to (19.center);
		\draw (19.center) to (18.center);
		\draw (15.center) to (16.center);
		\draw [style=Arrow] (14.center) to (13.center);
		\draw [style=Arrow] (10.center) to (12.center);
		\draw [style=Arrow] (26.center) to (27.center);
		\draw [style=Arrow, bend right=45, looseness=1.25] (16.center) to (29.center);
		\draw [bend right=45, looseness=1.25] (41.center) to (40.center);
		\draw [style=Arrow] (40.center) to (39.center);
		\draw (26.center) to (28.center);
		\draw [style=Arrow] (79.center) to (80.center);
		\draw [bend right=45, looseness=1.25] (28.center) to (74.center);
		\draw [style=Arrow] (74.center) to (86.center);
	\end{pgfonlayer}
\end{tikzpicture}}
\caption{Internal structure of the \IBIS package, which is described in the text. The processes with round corners are part of \IBIS while the other processes are not.}
\label{fig:ecosystem}
\end{figure}
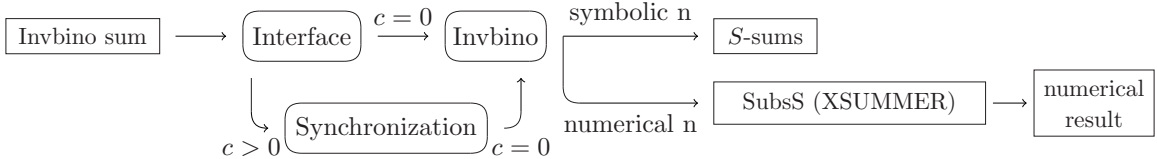

\noindent \textbf{Output} \\
The program is initiated by
\begin{equation}
\texttt{\#call IBIS}
\end{equation}
and solves inverse binomial sums in terms of $S$-sums. Internally, it follows the structure summarized in fig.~\ref{fig:ecosystem}: 
First the \texttt{Interface} of \IBIS checks whether the user's input of the inverse binomial sum follows the format that we specified in this section, and determines if synchronization of the denominator is necessary before solving (i.e. $c \neq 0$). If required, the package performs \texttt{Synchronization} prior to applying the \texttt{Invbino} procedure to solve the inverse binomial sum. If the user inputs a symbolic value for $n$, the resulting $S$-sums cannot be further reduced. However, if $n$ is numerical, the resulting $S$-sums can be simplified further using the \texttt{SubS} procedure from \XSUMMER, leading to a numerical result. 

\subsection{Examples}
\label{sec:examples}
To see \IBIS in action, we recall the sums given in the introduction, i.e.
\begin{align}
\text{\eqref{eq:exampleDouble}} &= 
-\sum_{k_2=1}^\infty \frac{1}{k_2(k_2+1)}\sum_{k_1=1}^{k_2-1}
(-1)^{k_1}\frac{k_1!(k_2-k_1)!}{k_2!}\frac{S_1(k_1)S_1(k_2-k_1)}{k_1^2}\,, 
\nn \\
\text{\eqref{eq:exampleTriple}} &= 
-\sum_{k_3=2}^\infty
\sum_{k_2=1}^{k_3-1} 
\frac{k_2! (k_3-k_2)!}{k_3!}
\frac{1}{(k_2+1)k_2(k_3-k_2)^2}
\sum_{k_1=1}^{k_2}
(-1)^{k_1}
\frac{k_1! (k_2-k_1)!}{k_2!}
\frac{1}{k_1^2}\,.
\nn \end{align}
For the double sum in \eq{exampleDouble}, the inner sum over $k_1$ can be performed with \IBIS,~yielding
\begin{align}
\sum_{k_1=1}^{k_2-1}(-1)^{k_1}
&\frac{k_1!(k_2-k_1)!}{k_2!}\frac{S_1(k_1)S_1(k_2-k_1)}{k_1^2}\nn\\
=&
- 3S_{-3,1}(k_2)
+ 3S_{1,-3}(k_2)
+ 2S_{1,-2,1}(k_2)
- 4S_{1,1,-2}(k_2)\nn\\
&- 2S_{1,1,2}(k_2)
- S_{1,2,1}(k_2)
+ 3S_{1,3}(k_2)
+ 4S_{2,-2}(k_2)
+ 2S_{2,2}(k_2)
\,.\end{align}
The final sum over $k_2$ to infinity can be performed with e.g.~\SUMMER, resulting in
\begin{equation}
\text{\eqref{eq:exampleDouble}} =\! 
\frac4{15}\log(2)^5 - \frac43\log(2)^3\zeta_2 +\! \frac72\log(2)^2\zeta_3 
+ 8\log(2)\textrm{Li}_4(\tfrac12) + 8\textrm{Li}_5(\tfrac12) 
- \frac32\zeta_2\zeta_3 - \frac{79}{16}\zeta_5,
\end{equation}
where we used the tables of \SUMMER to rewrite the multiple zeta values and Euler sums to the more familiar zeta and (poly)logarithmic values.

For the triple sum of \eq{exampleTriple}, we can perform both the sums over $k_1$ and $k_2$ with \IBIS. First, we note that the upper limit of the sum over $k_1$ is not yet in \IBIS form. One therefore splits off the upper limit $k_1=k_2$. Let us focus on the remaining sum, which can now be performed with \IBIS
\begin{equation}
\sum_{k_1=1}^{k_2-1}
(-1)^{k_1}
\frac{k_1! (k_2-k_1)!}{k_2!}
\frac{1}{k_1^2}
= 2S_{-2}(k_2)+ S_2(k_2)-\frac{(-1)^{k_2}}{k_2^2}\,.
\end{equation}
We notice that the last term cancels against the term $k_1=k_2$ we split off from the sum before. 
Now, before we can continue with \IBIS to perform the sum over $k_2$, one needs to split the denominators first
\begin{align}
\frac{1}{(k_2+1)k_2(k_3-k_2)^2} 
=& \frac{1}{k_2}\frac{1}{k_3^2}-\frac{1}{k_2+1}\frac{1}{(k_3+1)^2}
+\frac{1}{k_3-k_2}\left(\frac{1}{k_3^2}-\frac{1}{(k_3+1)^2}\right)\nn\\
&+\frac{1}{(k_3-k_2)^2}\left(\frac{1}{k_3}-\frac{1}{k_3+1}\right)
\,.\end{align}
The sums are now in the form of tab.~\ref{table:possible sums} and can therefore be performed with \IBIS. More details on how sums can be prepared as input for \IBIS can be found in the manual provided with the package. 

The full result of this sum is rather long, so we focus on a few terms. First of all, we consider 
\begin{align}
\sum_{k_2=1}^{k_3-1} 
\frac{k_2! (k_3-k_2)!}{k_3!}
\frac{S_{2}(k_2)}{k_2+1}
&=        
\frac{1}{(k_3+1)^3}
- \frac{S_2(k_3+1)}{k_3+1}
+(k_3+1)\biggl\{
2S_{1,1,1,1}(\tfrac12,2,\tfrac12,2;k_3\!+\!1)\nn\\
&- 4S_{1,1,2}(\tfrac12,2,1;k_3\!+\!1)
- 2S_{1,2,1}(\tfrac12,1,2;k_3\!+\!1)
+ 4S_{1,3}(\tfrac12,2;k_3\!+\!1)\nn\\
&- 2S_{2,1,1}(1,\tfrac12,2;k_3\!+\!1)
+ 5S_{2,2}(k_3\!+\!1)
+ 2S_{3,1}(\tfrac12,2;k_3\!+\!1)\nn\\
&- 5S_4(k_3\!+\!1)
\biggr\}\,,
\end{align}
which is a sum with a denominator shifted by $c=1$. The recursion that brings down $c$, explained in \sec{synch_denominator}, generates quite a few terms. 
Another example is
\begin{align}
\sum_{k_2=1}^{k_3-1} 
\frac{k_2! (k_3-k_2)!}{k_3!}
\frac{S_{2}(k_2)}{k_3-k_2}
&= \sum_{k_2=1}^{k_3-1} 
\frac{k_2! (k_3-k_2)!}{k_3!}
\frac{S_{2}(k_3-k_2)}{k_2} \nn\\
&=\Bigl(\frac12\Bigr)^{k_3}\biggl\{
- S_{1,1,1}(2,\tfrac12,2;k_3)
+ 2S_{1,2}(2,1;k_3)
+ S_{2,1}(1,2;k_3)
\nn \\ & 
- 2S_{3}(2;k_3)
\biggr\}\,,
\end{align}
and similar results for the other terms. While these expressions can be simplified by replacing single $S$-sums with products using \eq{$S$-sum algebra}, this is inconvenient for carrying out the remaining sum over $k_3$.
This remaining sum can now be performed with \XSUMMER. A \FORM program with the full answer is enclosed with the package.

%%%%%%%%%%%%%%%%%%%%%%%%%%%%%%
\subsection{Speed of the program}
\label{sec:speed}
%%%%%%%%%%%%%%%%%%%%%%%%%%%%%%
To demonstrate the performance improvement offered by \IBIS, we conducted a series of benchmark tests. We evaluated all inverse binomial sum types with low values for the extended weight that \IBIS can process without relying on external packages. These tests included:

\begin{itemize}
\item Inverse binomial sums without harmonic sums, for extended weights 1 through 6;
\item Inverse binomial sums with one harmonic sum, for weights 1 through 5;
\item Inverse binomial sums with two harmonic sums, for weights 2 through 5.
\end{itemize}
We compare \IBIS with \SIGMA/\textsc{EvaluateMultiSums}.

For each extended weight, we determined the average time required to evaluate the corresponding inverse binomial sums. For low weights, all possible sums of that weight were included in the benchmark. For higher extended weights, however, the total number of distinct inverse binomial sums becomes very large. Since evaluating each of these with \SIGMA would require prohibitively long runtimes, we instead used a random sample of the possible sums when benchmarking \SIGMA, while \IBIS was still applied to the full set. In both cases, the reported values represent the average evaluation time per sum for that extended weight. All tests were run on a 13th Gen Intel i7-13700H, 2400 MHz. 

\begin{figure}[!t]
\centering
    \begin{subfigure}[b]{0.45\textwidth}
        \centering
        \includegraphics[width=\linewidth]{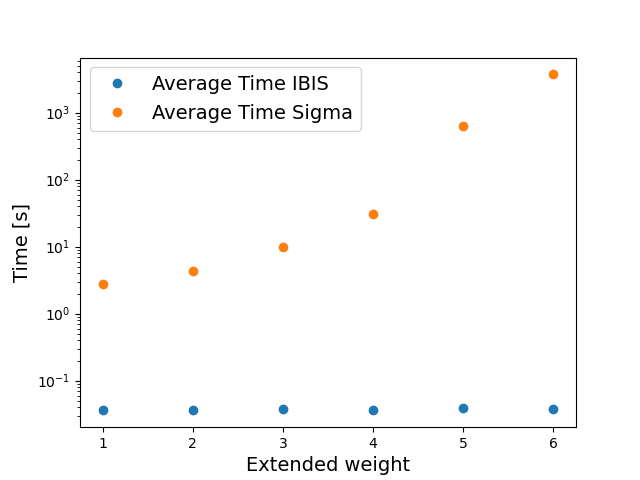}
        \caption{$\sum_{j=1}^{n-1} \frac{j!(n-j)!}{n!} \frac{1}{j^k}$}\label{fig:0sspeed}
    \end{subfigure}
    \begin{subfigure}[b]{0.45\textwidth}
        \centering
        \includegraphics[width=\linewidth]{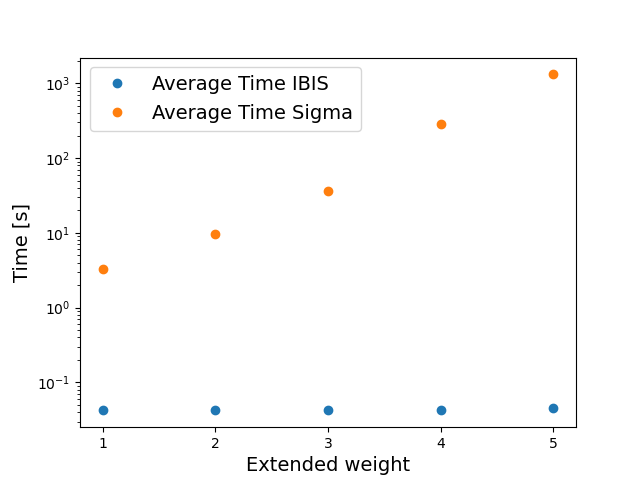}
        \caption{$ \sum_{j=1}^{n-1} \frac{j!(n-j)!}{n!} \frac{(-1)^j}{j^k}S_{p_1,\dots,p_k}(j)$}\label{fig:lowerindicesspeed}
    \end{subfigure} \\
    \begin{subfigure}[b]{0.45\textwidth}
        \centering
        \includegraphics[width=\linewidth]{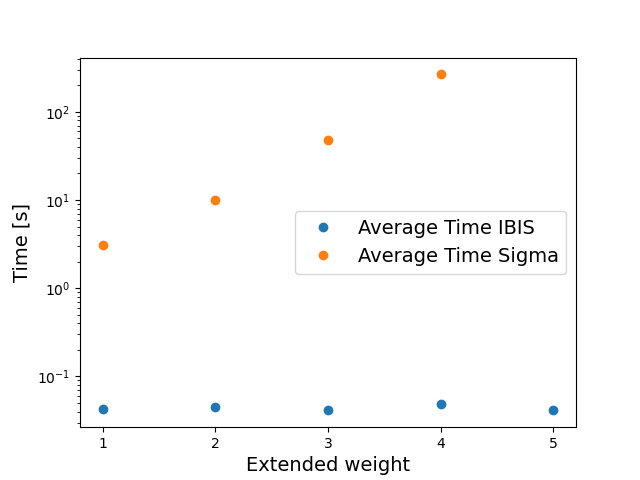}
        \caption{$\sum_{j=1}^{n-1} \frac{j!(n-j)!}{n!} \frac{1}{j^k}S_{p_1,\dots,p_k}(n-j)$}\label{fig:upperindicesspeed}
    \end{subfigure}     
    \begin{subfigure}[b]{0.45\textwidth}%
        \centering
        \includegraphics[width=\linewidth]{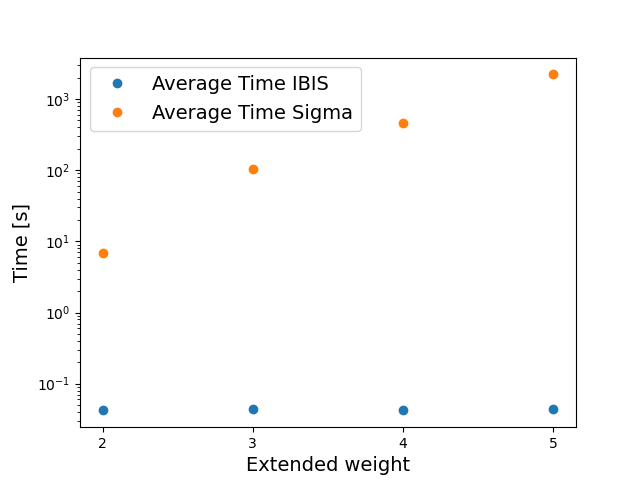}
        \caption{$\sum_{j=1}^{n-1} \frac{j!(n-j)!}{n!} \frac{1}{j^k}S_{p_1,\dots,p_k}(n-j)S_{q_1,\dots,q_r}(j)$} \label{fig:doubleSspeed}
    \end{subfigure}
    \caption{Plots showing the average execution times for evaluating inverse binomial sums of various extended weights using \IBIS and \SIGMA/\textsc{EvaluateMultiSums}. The top left plot presents results for inverse binomial sums without harmonic sums. The top right shows the results that include a single harmonic sum with $j$ as the argument, while the bottom left figure is for the case of one harmonic sum with argument $n-j$. The bottom right plot displays results for inverse binomial sums containing two harmonic sums. The expressions shown in the subcaptions are representative examples of these types of sums, though not exhaustive. For a complete list, see Table~\ref{table:possible sums}.} \label{fig:speedtests}
\end{figure}

The results are shown in \fig{speedtests}, with each panel corresponding to a different class of inverse binomial sums, with 0, 1, or 2 harmonic sums. \IBIS exhibits consistently low execution times across all extended weights in the test cases. This performance stems from the fact that \IBIS is specifically optimized for the particular class of inverse binomial sums used in the benchmark, with all necessary recurrence relations pre-derived and hardcoded into the system. 

\IBIS{}’s runtimes scale exponentially with extended weight, which can be attributed to the nature of the implemented recursions. Due to fraction splitting, the recursions involve sums that extend up to e.g.~the initial exponent or the value of one of the indices of the inverse binomial sum, as seen in \eq{finalrec}. Thus, increasing the weight or indices produces a significantly higher number of terms for \FORM to process, naturally slowing down computation.

This exponential runtime scaling is not visible for \IBIS in this figure. Indeed, for the relatively low-weight sums tested here, \IBIS{}’s runtimes are largely dominated by consistent \FORM startup overhead rather than the intrinsic complexity of the sums. To investigate this, we ran a loop solving the same inverse binomial sum (containing one harmonic sum of weight 2) 100 times in the same \FORM program. Since the fastest time for these sums is around 0.03 seconds, we expected a minimum of 3 seconds for the test. However, the program completed the task in just 0.48 seconds, indicating that solving low-weight sums doesn’t require enough time to increase the average time substantially. Indeed, this is also responsible for the small fluctuations seen in the time used by \IBIS.

Comparing to \SIGMA, \IBIS is orders of magnitude faster than \SIGMA. This is not surprising since \SIGMA is designed to handle a much broader range of sums, and must symbolically derive appropriate recursions on the fly for each case. This generality comes at the cost of significantly higher computation time.

In summary, while both \IBIS and \SIGMA exhibit exponential growth in execution time, \IBIS{}’s specialization and precompiled recursions give it a clear advantage for the types of sums studied here. This makes it particularly effective for high-performance symbolic evaluation in focused applications, like the example in App.~\ref{app:example}.

In sec.~\ref{sec:Recursionsec}, we also discussed the recursion implemented to synchronize the denominator of the inverse binomial sum with the argument of the harmonic sum in \eq{ctoc-1final}. This synchronization was performed one step at a time (i.e.~$c \to c-1$), generating numerous additional terms for \FORM to handle. Therefore, it’s important to note that attempting to solve inverse binomial sums with an unsynchronized denominator can significantly increase the runtime, beyond the cases shown in \fig{speedtests}.

%%%%%%%%%%%%%%%%%%%%%%%%%%%%%%%%
\section{Conclusion and outlook}
\label{sec:conclusion}
%%%%%%%%%%%%%%%%%%%%%%%%%%%%%%%%
In this paper, we presented \IBIS, a practical tool designed to solve certain classes of inverse binomial sums that arise in multi-loop calculations. We have chosen \FORM for its implementation, because of its speed, efficiency and capability to handle large symbolic expressions. 
Indeed, the sums we discuss can be handled by \SIGMA and \textsc{EvaluateMultiSums}, but our package is much faster for the dedicated set of inverse binomial sums we focus on in this paper.
\IBIS employs recursions to express inverse binomial sums in terms of $S$-sums, maintaining an analytic dependence on the upper limit. We explained in detail how these recursions are constructed using telescoping series. Furthermore, we addressed potential complications, such as the synchronization of  elements within the inverse binomial sums. To illustrate the relevance of inverse binomial sums, we provided an example of a loop calculation where these sums naturally appear in the appendix. Moreover, we demonstrated that the program is capable of solving inverse binomial sums of weights up to 6 relatively  quickly. By offering this tool, we aim to broaden the scope of possible loop calculations, thereby contributing to advancements in high-order perturbative computations. The GitHub repository includes a detailed manual and sample calculations alongside \IBIS.

While \IBIS successfully tackles the class of inverse binomial sums it was designed to solve, there are also sums that involve a combination of binomial and inverse binomial sums that it can't handle, such as in \eq{finalexampleIBIS}.
This sum, or the binomial part of it, can be carried out using \SIGMA and \textsc{EvaluateMultiSums}, solving recursions on the fly. Furthermore, certain classes of binomial sums are solved within \SUMMER and for others one can use a similar approach to that employed here. This should result in a similar speedup over on-the-fly recursion solving as seen in sec.~\ref{sec:speed}.

In conclusion, the development of \IBIS marks a step forward in efficiently
solving inverse binomial sums that appear in multi-loop calculations, contributing to the analytical toolkit available for high-order calculations in quantum field theory.

\acknowledgments
We would like to thank J.A.M. Vermaseren for his assistance in establishing the first telescoping series for solving inverse binomial sums and for valuable discussions. We thank J.~Bl\"umlein for pointing out an issue in our example and directing us to relevant literature.

\appendix
%%%%%%%%%%%%%%%%%%%%%%%%%%%%%%
\section{Example}
\label{app:example}
%%%%%%%%%%%%%%%%%%%%%%%%%%%%%%

\begin{figure}
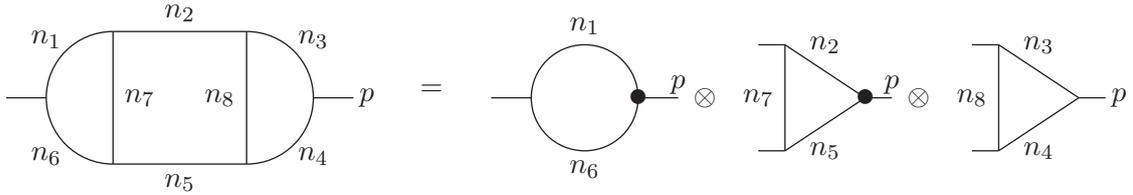

    \centering
    \subfloat{
        \begin{axopicture}(130,70)(-65,-35)
    		\Line(-65,0)(-50,0)
    		\Arc(-25,0)(25,90,270)
    		\Line(-25,25)(25,25)
    		\Line(-25,-25)(25,-25)
    		\Arc(25,0)(25,270,90)
    		\Line(-25,-25)(-25,25)
    		\Line(25,-25)(25,25)
			\Line(50,0)(65,0)
    		\Text(-50,22){$n_1$}
    		\Text(0,32){$n_2$}
    		\Text(50,22){$n_3$}
    		\Text(50,-22){$n_4$}
    		\Text(0,-32){$n_5$}
    		\Text(-50,-22){$n_6$}
    		\Text(-15,0){$n_7$}
    		\Text(15,0){$n_8$}
            \Text(70,0){$p$}
    	\end{axopicture}
    }
	\hspace{0.75cm}
    \raisebox{9\height}{$=$}
    \hspace{0.25cm}
    \subfloat{
        \begin{axopicture}(230,70)(0,-35)
    		\Line(0,0)(15,0)
    		\Arc(35,0)(20,0,360)
			\Line(55,0)(70,0)
    		\Text(35,27){$n_1$}
			\Text(35,-27){$n_6$}
            \Text(70,5){$p$}
            \Text(55,0){\Large$\bullet$}
			\Text(80,0){$\otimes$}
			\Line(100,20)(110,20)
			\Line(100,-20)(110,-20)
			\Line(110,20)(110,-20)
			\Line(110,20)(140,0)
			\Line(110,-20)(140,0)
			\Line(140,0)(150,0)
            \Text(140,0){\Large$\bullet$}
			\Text(125,20){$n_2$}
			\Text(125,-20){$n_5$}
			\Text(100,0){$n_7$}
            \Text(150,5){$p$}
			\Text(160,0){$\otimes$}
			\Line(180,20)(190,20)
			\Line(180,-20)(190,-20)
			\Line(190,20)(190,-20)
			\Line(190,20)(220,0)
			\Line(190,-20)(220,0)
			\Line(220,0)(230,0)
			\Text(205,20){$n_3$}
			\Text(205,-20){$n_4$}
			\Text(180,0){$n_8$}
            \Text(235,0){$p$}
    	\end{axopicture}
    }
    \caption{Factorization in terms of diagrams. The three-loop two-point function on the left-hand side can be expressed as the sequential insertion of simpler components: a one-loop three-point function inserted into another one-loop three-point function, which is then inserted into a one-loop two-point function. The insertion points are located at the dark vertex.}
    \label{fig:ThreeLoopLadder}
\end{figure}

To illustrate an actual calculation of a Feynman integral that results in inverse binomial sums solvable by \IBIS, consider the three-loop two-point integral depicted on the left-hand side of Fig.~\ref{fig:ThreeLoopLadder}. We employ dimensional regularization with $d=4-2\epsilon$, which preserves gauge invariance and facilitates the evaluation of the integral corresponding to the three-loop ladder diagram 
\begin{align} \label{eq:ladderdef}
\hat{I}^{(3,8)}(n_1,\ldots,n_8) 
&= c_{\Gamma}^{-3} (-p^2)^{n_{12345678}-3(2-\epsilon)} \int \frac{\df^dk_1}{\text{i}\pi^{d/2}}\int \frac{\df^dk_2}{\text{i}\pi^{d/2}}\int \frac{\df^dk_3}{\text{i}\pi^{d/2}}
\prod_{j=1}^8 \frac{1}{(-k_j^2)^{n_j}}\,,
\end{align}
where $p$ is the external momentum, the momenta of the other propagators in the ladder diagram are $k_4=k_3-p$, $k_5=k_2-p$, $k_6=k_1-p$, $k_7=k_2-k_1$, $k_8=k_3-k_2$ and the constant $c_\Gamma$ is given by
\begin{align}
     c_\Gamma = \frac{\Gamma(1+\epsilon)\Gamma(1-\epsilon)^2}{\Gamma(1-2\epsilon)} \,.
\end{align}
The superscript in $\hat{I}^{(l,m)}$ indicates the number of loops $l$ and the number of propagators $m$.\footnote{We include the hat to follow the notation of ref.~\cite{Bierenbaum:2003ud} indicating that the mass scales have been factored out.}
The prefactor ensures that $\hat{I}^{(3,8)}$ is independent of $(-p^2)$ and avoids the repeated appearance of $\gamma_E$. We employ the shorthand notation $n_{ab \cdots } = n_{a} + n_{b} + \dots $ to denote the sum of indices, where $a,b,\dots, \in \{ 1,2,...,8 \}$.

In the remainder of this appendix we will show that the three-loop integral in \eq{ladderdef} can be expressed as a convolution product of three one-loop integrals~\cite{Bierenbaum:2003ud}. To that end, we first introduce the following one-loop integrals
\begin{align} \label{eq:oneloopdef}
\hat{I}^{(1,2)}(n_1,n_6) 
&= c_\Gamma^{-1} (-p^2)^{n_{16}-2+\epsilon} \int \frac{\df^dk_1}{\text{i}\pi^{d/2}} \frac{1}{(-k_1^2)^{n_1}(-k_6^2)^{n_6}} \,,\nn\\
I^{(1,3)}(n_2,n_5,n_7;x,y) 
&= c_\Gamma^{-1} (-p^2)^{n_{257}-2+\epsilon} \int \frac{\df^dk_2}{\text{i}\pi^{d/2}} \frac{1}{(-k_2^2)^{n_2}(-k_5^2)^{n_5}(-k_7^2)^{n_7}} \,, \nn \\
I^{(1,3)}(n_3,n_4,n_8;w,z) 
&= c_\Gamma^{-1} (-p^2)^{n_{348}-2+\epsilon} \int \frac{\df^dk_3}{\text{i}\pi^{d/2}} \frac{1}{(-k_3^2)^{n_3}(-k_4^2)^{n_4}(-k_8^2)^{n_8}} \,,
\end{align}
where $x = p^2/k_1^2$, $y= p^2/k_6^2$, $w = p^2/k_2^2$ and $z = p^2/k_5^2$. The one-loop bubble integral, $\hat{I}^{(1,2)}(n_1,n_6)$, can be evaluated as
\begin{align}
\hat{I}^{(1,2)}(n_1,n_6)
&= c_\Gamma^{-1} \frac{\Ga(n_{16}+\eps-2)}{\Ga(n_1)\Ga(n_6)}\frac{\Ga(2-\eps-n_1)\Ga(2-\eps-n_6)}{\Ga(4-2\eps-n_{16})} \,.
\end{align}
The two triangle integrals can be represented as double MB integrals
\begin{align}
I^{(1,3)}(n_2,n_5,n_7;x,y) 
&= \frac{1}{(2\pi \text{i})^2}
\int_{\gamma_1-\text{i}\infty}^{\gamma_1+\text{i}\infty}\df\sigma
\int_{\gamma_2-\text{i}\infty}^{\gamma_2+\text{i}\infty}\df\tau
\,x^{-\sigma}
y^{-\tau}\,
\hat{I}^{(1,3)}(n_2,n_5,n_7;\tau,\sigma) \,, \nn \\
I^{(1,3)}(n_3,n_4,n_8;w,z) 
&= \frac{1}{(2\pi \text{i})^2}
\int_{\gamma_3-\text{i}\infty}^{\gamma_3+\text{i}\infty}\df\sigma
\int_{\gamma_4-\text{i}\infty}^{\gamma_4+\text{i}\infty}\df\tau \,
w^{-\sigma}
z^{-\tau} \,
\hat{I}^{(1,3)}(n_3,n_4,n_8;\tau,\sigma) \,.\label{eq:MBrepOneLoopTriangle}
\end{align}
Here, the functions $\hat{I}^{(1,3)}(n_2,n_5,n_7;\tau,\sigma)$ and $\hat{I}^{(1,3)}(n_3,n_4,n_8;\tau,\sigma)$ are given by
\begin{align}
\hat{I}^{(1,3)}(n_2,n_5,n_7;\tau,\sigma)
=& \, c_\Gamma^{-1} \frac{1}{\Ga(n_2)\Ga(n_5)\Ga(n_7)\Ga(4-2\eps-n_{257})}\nn\\
&\times
\Ga(-\sigma)\Ga(-\sigma+2-\eps-n_{57})\Ga(-\tau)\Ga(-\tau+2-\eps-n_{27})\nn\\
&\times
\Ga(\sigma+\tau+n_{257}+\eps-2)\Ga(\sigma+\tau+n_7) \,, \nn \\
\hat{I}^{(1,3)}(n_3,n_4,n_8;\tau,\sigma)
=& \,c_\Gamma^{-1} \frac{1}{\Ga(n_3)\Ga(n_4)\Ga(n_8)\Ga(4-2\eps-n_{348})}\nn\\
&\times
\Ga(-\sigma)\Ga(-\sigma+2-\eps-n_{48})\Ga(-\tau)\Ga(-\tau+2-\eps-n_{38})\nn\\
&\times
\Ga(\sigma+\tau+n_{348}+\eps-2)\Ga(\sigma+\tau+n_8) \,.
\end{align}
Although these two definitions merely swap symbols, providing the explicit expression makes further substitution easier.
The integration contours in \eq{MBrepOneLoopTriangle} run parallel to the imaginary axis and are chosen in such a way that the poles of $\Ga(-\sigma+\ldots)$, $\Ga(-\tau+\ldots)$ and $\Ga(-\sigma-\tau+\ldots)$ are to the right of the contour and the poles of $\Ga(\sigma+\ldots)$, $\Ga(\tau+\ldots)$ and $\Ga(\sigma+\tau+\ldots)$ are to the left of the contour. 

We are now in the position to show that the three-loop integral in \eq{ladderdef} can be represented as a convolution product. First, we rewrite the expression for the three-loop ladder diagram in \eq{ladderdef} as follows
\begin{align}
&\hat{I}^{(3,8)}(n_1,\ldots,n_8) \nn \\
&= c_{\Gamma}^{-3} (-p^2)^{n_{12345678}-3(2-\epsilon)} \int \frac{\df^dk_1}{\text{i}\pi^{d/2}} \frac{1}{(-k_1^2)^{n_1}(-k_6^2)^{n_6}}\int \frac{\df^dk_2}{\text{i}\pi^{d/2}} \frac{1}{(-k_2^2)^{n_2}(-k_5^2)^{n_5}(-k_7^2)^{n_7}} \nn \\
&\hspace{1cm} \times \int \frac{\df^dk_3}{\text{i}\pi^{d/2}} \frac{1}{(-k_3^2)^{n_3}(-k_4^2)^{n_4}(-k_8^2)^{n_8}}\,.
\end{align}
The integral over $k_3$ is identified with the triangle integral on the third line of \eq{oneloopdef}, so we substitute its double MB representation from \eq{MBrepOneLoopTriangle}
\begin{align}
&\hat{I}^{(3,8)}(n_1,\ldots,n_8) \nn \\
&= c_{\Gamma}^{-2} (-p^2)^{n_{12567}-2(2-\epsilon)} \frac{1}{(2\pi \text{i})^2} \int_{\gamma_3-\text{i}\infty}^{\gamma_3+\text{i}\infty}\df\sigma_2
\int_{\gamma_4-\text{i}\infty}^{\gamma_4+\text{i}\infty}\df\tau_2 \,\int \frac{\df^dk_1}{\text{i}\pi^{d/2}} \frac{1}{(-k_1^2)^{n_1}(-k_6^2)^{n_6}} \nn \\
&\hspace{1cm} \times \int \frac{\df^dk_2}{\text{i}\pi^{d/2}} \frac{(-p^2)^{-\sigma_2}(-p^2)^{-\tau_2}}{(-k_2^2)^{n_2-\sigma_2}(-k_5^2)^{n_5-\tau_2}(-k_7^2)^{n_7}}
\hat{I}^{(1,3)}(n_3,n_4,n_8;\tau_2,\sigma_2)\,.
\end{align}
Next, the integral over $k_2$ is identified with the triangle integral on the second line of \eq{oneloopdef}. Substituting its MB representation yields
\begin{align}
&\hat{I}^{(3,8)}(n_1,\ldots,n_8) \\
&= c_{\Gamma}^{-1} (-p^2)^{n_{16}-(2-\epsilon)} \frac{1}{(2\pi \text{i})^4}
\int_{\gamma_1-\text{i}\infty}^{\gamma_1+\text{i}\infty}\df\sigma_1
\int_{\gamma_2-\text{i}\infty}^{\gamma_2+\text{i}\infty}\df\tau_1
\int_{\gamma_3-\text{i}\infty}^{\gamma_3+\text{i}\infty}\df\si_2
\int_{\gamma_4-\text{i}\infty}^{\gamma_4+\text{i}\infty}\df\tau_2 \int \frac{\df^dk_1}{\text{i}\pi^{d/2}} \nn\\
&\hspace{1cm}\times\frac{(-p^2)^{-\sigma_1}(-p^2)^{-\tau_1}}{(-k_1^2)^{n_1-\sigma_1}(-k_6^2)^{n_6-\tau_1}}
\hat{I}^{(1,3)}(n_2-\sigma_2,n_5-\tau_2,n_7;\tau_1,\sigma_1)\hat{I}^{(1,3)}(n_3,n_4,n_8;\tau_2,\sigma_2) \,. \nn
\end{align}
Finally, we recognize  the remaining integral over $k_1$ as the bubble integral on the first line of \eq{oneloopdef}. Thus, we obtain the quadruple MB representation for the three-loop ladder diagram
\begin{align} 
&\hat{I}^{(3,8)}(n_1,\ldots,n_8) \nn \\
&= \frac{1}{(2\pi \text{i})^4}
\int_{\gamma_1-\text{i}\infty}^{\gamma_1+\text{i}\infty}\df\sigma_1
\int_{\gamma_2-\text{i}\infty}^{\gamma_2+\text{i}\infty}\df\tau_1
\int_{\gamma_3-\text{i}\infty}^{\gamma_3+\text{i}\infty}\df\si_2
\int_{\gamma_4-\text{i}\infty}^{\gamma_4+\text{i}\infty}\df\tau_2\
\hat{I}^{(1,2)}(n_1-\sigma_1,n_6-\tau_1)\nn\\
&\hspace{1cm}\times \hat{I}^{(1,3)}(n_2-\sigma_2,n_5-\tau_2,n_7;\tau_1,\sigma_1)
\hat{I}^{(1,3)}(n_3,n_4,n_8;\tau_2,\sigma_2) \,.\label{eq:MBrepThreeLoopLadder}
\end{align}
This final representation corresponds diagrammatically to the structure shown on the right-hand side of Fig.~\ref{fig:ThreeLoopLadder}. In this picture the product structure reflects the insertion of one diagram into another.

In order to proceed with \eq{MBrepThreeLoopLadder}, we use the \texttt{Mathematica} packages \texttt{MB.m} and \texttt{MBsums.m} to close the MB contours. In this case, the process results in nested sums with a maximum depth of four. These resulting sums generally share a similar structure. To illustrate this, consider one such residue sum obtained from closing the contour, corresponding to \eq{exampleDouble},
\begin{equation}
\sum_{k_1=0}^\infty \sum_{k_2=0}^\infty 
(-1)^{k_1}\frac{k_1!k_2!}{(k_1+k_2+2)!}\frac{S_1(k_1+1)S_1(k_2)}{(k_1+k_2+1)(k_1+1)}
\,.
\end{equation}
In order to see that this double sum indeed contains an inverse binomial sum, we apply the substitution $k_2'= k_1+k_2$. This yields 
\begin{align}
\sum_{k_1=0}^\infty 
&\sum_{k_2'=k_1}^\infty
(-1)^{k_1}\frac{k_1!(k_2'-k_1)!}{(k_2'+2)!}\frac{S_1(k_1+1)S_1(k_2'-k_1)}{(k_2'+1)(k_1+1)}\nn\\
&= \sum_{k_2=0}^\infty \frac{1}{(k_2+2)(k_2+1)^2}\sum_{k_1=0}^{k_2}
(-1)^{k_1}\frac{k_1!(k_2-k_1)!}{k_2!}\frac{S_1(k_1+1)S_1(k_2-k_1)}{(k_1+1)}
\end{align}
where in the second line we switched the sums around and dropped the prime on $k_2$. Next, shifting $k_1\to k_1+1$ and $k_2\to k_2+1$ we get
\begin{equation}
-\sum_{k_2=1}^\infty \frac{1}{k_2(k_2+1)}\sum_{k_1=1}^{k_2-1}
(-1)^{k_1}\frac{k_1!(k_2-k_1)!}{k_2!}\frac{S_1(k_1)S_1(k_2-k_1)}{k_1^2}\,.
\end{equation}
Note that the upper limit of the inner sum is formally $k_2$, but the term at $k_1=k_2$ vanishes as $S_1(k_2-k_1) = 0$. In the notation of tab.~\ref{table:possible sums}, we see that the inner sum is a sign-altering inverse binomial sum with two S-sums.  

As a second example, in \eq{exampleTriple}, we consider another residue sum
\begin{equation}
\sum_{k_1=0}^\infty \sum_{k_2=0}^\infty \sum_{k_3=0}^\infty
(-1)^{k_1}
\frac{k_1! k_2!k_3!}{(k_1\!+\!k_2\!+\!k_3\!+\!2)!}
\frac{1}{(k_1\!+\!1)(k_1\!+\!k_2\!+\!2)(k_1\!+\!k_2\!+\!1)(k_3\!+\!1)}\,.
\end{equation}
By successively applying the substitution $k_2'= k_1+k_2$ and $k_3'= k_2'+k_3$ we obtain
\begin{equation}
\sum_{k_1=0}^\infty \sum_{k_2'=k_1}^\infty \sum_{k_3'=k_2'}^\infty
(-1)^{k_1}
\frac{k_1! (k_2'-k_1)!(k_3'-k_2')!}{(k_3'\!+\!2)!}
\frac{1}{(k_1\!+\!1)(k_2'\!+\!2)(k_2'\!+\!1)(k_3'\!-\!k_2'\!+\!1)}
\,.
\end{equation}
Dropping the primes on the summation variables and multiplying the numerator and denominator by $k_2!$ yields
\begin{equation}
\sum_{k_3=0}^\infty
\sum_{k_2=0}^{k_3} 
\frac{k_2! (k_3-k_2)!}{(k_3+2)!}
\frac{1}{(k_2\!+\!2)(k_2\!+\!1)(k_3\!-\!k_2\!+\!1)}
\sum_{k_1=0}^{k_2}
(-1)^{k_1}
\frac{k_1! (k_2-k_1)!}{k_2!}
\frac{1}{(k_1\!+\!1)}
\,,
\end{equation}
where we interchanged the order of summation. 
By shifting the summation variables as $k_1\to k_1+1$, $k_2\to k_2+1$ and $k_3\to k_3+2$ we can (almost) express these sums in the notation of tab.~\ref{table:possible sums}:
\begin{equation}
-\sum_{k_3=2}^\infty
\sum_{k_2=1}^{k_3-1} 
\frac{k_2! (k_3-k_2)!}{k_3!}
\frac{1}{(k_2+1)k_2(k_3-k_2)^2}
\sum_{k_1=1}^{k_2}
(-1)^{k_1}
\frac{k_1! (k_2-k_1)!}{k_2!}
\frac{1}{k_1^2}
\,.
\end{equation}
The inner sum over $k_1$ (up to the upper bound) is a sign-altering inverse binomial sum, whereas the middle sum over $k_2$ is a fixed sign inverse binomial sum. 

A final sum we consider is
\begin{equation}
\sum_{k_1=1}^\infty \sum_{k_2=1}^\infty \sum_{k_3=1}^\infty
\frac{(k_1+k_2+1)!(k_1+k_3)! S_2(k_3)}{(k_2+1)(k_3+1)k_1!(k_1+k_2+k_3+3)!}
\,,\end{equation}
which was presented in \eq{finalexampleIBIS}.
By successively applying the substitution $k_3'= k_1+k_3$ and $k_2'= k_2+k_3'$ we obtain
\begin{equation}
\sum_{k_1=1}^\infty \sum_{k_3'=k_1+1}^\infty \sum_{k_2'=k_3'+1}^\infty
\frac{(k_1+k_2'-k_3'+1)!\,k_3'!\, S_2(k_3'-k_1)}{(k_2'-k_3'+1)(k_3'-k_1+1)k_1!(k_2'+3)!}
\,.\end{equation}
Similar to the previous example, we can drop the primes on the summation variables $k_2$ and $k_3$ and multiply the numerator and the denominator now by $(k_2-k_3)!$
\begin{equation}
\sum_{k_2=3}^\infty 
\sum_{k_3=2}^{k_2-1} \frac{k_3!(k_2-k_3)!}{(k_2+3)!}
\sum_{k_1=1}^{k_3-1} \frac{(k_1+k_2-k_3+1)!}{k_1!(k_2-k_3+1)!}\frac{S_2(k_3-k_1)}{k_3-k_1+1}\,,
\end{equation}
where we interchanged the order of summation. 
In the notation of Tab.~\ref{table:possible sums}, this sum is expressed as
\begin{equation}
\sum_{k_2=3}^\infty \frac{1}{(k_2+3)(k_2+2)(k_2+1)}
\sum_{k_3=2}^{k_2-1} \frac{k_3!(k_2-k_3)!}{k_2!}
\sum_{k_1=1}^{k_3-1} \binom{k_1\!+\!k_2\!-\!k_3\!+\!1}{k_1}\frac{S_2(k_3-k_1)}{k_3-k_1+1}.
\end{equation}

%----------------------------------------------------------------------------------------
%	REFERENCE LIST
%----------------------------------------------------------------------------------------
\bibliographystyle{JHEP}
% \bibliography{BibfileIBIS}

\providecommand{\href}[2]{#2}\begingroup\raggedright\endgroup

\end{document}